# Lithography-free, apodized thickness-shear $LiNbO_3$ resonators for piezoelectric DC-DC power conversion

Xu Zhao[1], Shun Yao[1] and Roy H. Olsson III[1]*

[1]Department of Electrical and Systems Engineering, University of Pennsylvania, Philadelphia, PA, USA

*Corresponding author Email: rolsson@engineering.upenn.edu

## Abstract

Piezoelectric resonators have emerged as promising energy-storage elements for compact high-frequency DC-DC power converters. Still, practical implementation requires simultaneously high $k_t^2$ (electromechanical coupling), $Q$ (high quality factor) and low resistance/impedance fluctuations. Spurious Lamb-wave and lateral modes are particularly problematic because they reduce the usable inductive frequency span and complicate converter operation over variable loads. Here we report a lithography-free, apodized 163° Y-cut lithium niobate thickness-shear resonator with reduced electrically visible spurious responses while preserving the intrinsically large coupling of the thickness-shear mode. Finite-element simulations show that a square plate excites numerous lateral spurious modes near the principal shear resonance. In contrast, a curved asymmetric plate outline redistributes the lateral acoustic boundary conditions and reduces electrically visible spurious responses. Experiments on multiple resonator outlines, interconnection schemes, and device areas are performed to characterize the device responses. The representative 560 $\mathrm{mm}^2$ device exhibits a series resonance near 3.67 MHz, an effective electromechanical coupling coefficient of 53.57%, a series quality factor of 1146, and a series resonance figure of merit of 614. A representative device fitted with a modified Butterworth-van Dyke equivalent circuit model shows excellent agreement between measurement and model, with a maximum Bode quality factor of 2896. These results support lateral-boundary apodization as a practical approach to high-coupling lithium niobate resonators with reduced electrically visible spurious responses for piezoelectric power converters.

## Introduction

Achieving higher power density in DC-DC converters critically depends on reducing the footprint of passive energy-storage components. Although increasing the switching frequency enables smaller magnetic inductors and transformers, further miniaturization is limited by frequency-dependent core and winding losses, reduced effective permeability, parasitic resistance, and practical thermal constraints[1–3]. Piezoelectric resonators (PRs) have therefore emerged as an alternative energy-storage platform in which electrical energy is reversibly converted to acoustic energy through a high-$Q$ mechanical resonance[4–7]. System-level demonstrations and circuit studies have shown that PR-based converters can reach very high resonator power density, including a 1.01 kW/cm³ resonator power-density demonstration, and can support compact magnetic-less or magnetics-reduced converter architectures[8–11]. At the same time, recent studies of nonlinear losses and material limits show that resonator-level performance, thermal management, and spectral purity remain central barriers to practical high-power operation[12].

In piezoelectric resonator based power converters, $k_t^2$ and $Q$ are critical performance metrics because they directly influence the achievable conversion ratio, load range, and resonator-limited efficiency[7,13]. Lithium niobate ($LiNbO_3$) is particularly attractive for piezoelectric power conversion because it combines strong piezoelectric response, low acoustic loss and wafer-scale availability[14,15]. For a converter whose efficiency is limited by the piezoelectric resonator, the theoretical maximum efficiency can be expressed in a form that decreases with the inverse electromechanical figure of merit ($k_t^2Q$), as shown in Eq. (1)[7]:

$$\eta_{max} = \frac{1}{1 + \frac{\pi^3}{8k_t^2 Q}\frac{V_{in} + V_{out}}{V_{out}}} \tag{1}$$

This relation emphasizes why resonators for power conversion must achieve large $k_t^2Q$ products rather than optimizing coupling or $Q$ alone. However, high coupling and high $Q$ are insufficient if the resonance spectrum contains strong spurious modes. In PR-based converters, useful operation often occurs in the inductive region between the series and parallel resonances, and spurious modes within or near this region can interrupt the monotonic impedance trajectory, reduce the accessible load range, and force fixed-frequency controllers away from the most efficient operating point[9]. Importantly, a smooth admittance magnitude alone does not guarantee a suitable response for power conversion; narrow spurious modes can produce fluctuations in the real part of the impedance, R, and thereby increase resonator loss near the intended operating region. Consequently, recent LN (lithium niobate) resonator research has increasingly focused on suppressing spurious modes while retaining high coupling and low motional resistance.

Several design strategies have been developed to address this challenge. Braun et al. reported optimized LN resonators for piezoelectric power conversion, but the response was constrained by pronounced spurious features[7]. Nguyen et al. subsequently demonstrated a near-spurious-free $LiNbO_3$ thickness-shear (TS) resonator with $Q_{max}$ of approximately 3500 and $k_t^2$ of 45% at 5.94 MHz by selecting a favorable crystal orientation and using a busbar-free acoustic design[16]. For $LiNbO_3$ thickness-extensional (TE) devices, grounded-ring and active-ring concepts have also been used to suppress spurious modes, with reported TE-LN resonators achieving $k_t^2$ near 30% and $Q$ values up to approximately 4000[17]. More recently, Yao et al. demonstrated twist piezoelectricity in inversely stacked and twisted bilayer X-cut $LiNbO_3$, achieving a high $k_t^2$ of 85.5% through twist-angle-controlled polarization[18]. Their work also showed that single-layer X-cut $LiNbO_3$ supports a secondary ($\beta$)-polarized mode that appears as a pronounced spurious peak in the admittance response. However, the bilayer twist approach requires precise wafer bonding and angular alignment, increasing fabrication complexity compared with single-layer resonator designs. Beyond single-layer LN, lithium tantalate resonators and periodically poled piezoelectric-film LN structures have been explored to improve power-conversion frequency scaling and figure of merit[19,20]. These advances show rapid progress in acoustic resonator design for power conversion, but they mainly rely on electrode patterning, crystal-orientation selection, or multilayer polarization engineering[21,22].

TS modes in $LiNbO_3$ remain especially compelling because they can provide very large coupling at megahertz frequencies while retaining a mechanically robust bulk-plate geometry. Nevertheless, plate-shaped TS resonators are vulnerable to lateral standing waves and Lamb-wave-like parasitic modes generated by finite lateral dimensions and discontinuous electrical or mechanical boundary conditions. The present work applies the established principle of apodization to the lateral plate boundary of the resonator. We demonstrate a lithography-free, apodized TS-mode $LiNbO_3$ resonator in which the resonator outline is defined by laser machining after blanket electrode deposition. A same-area COMSOL comparison and measurements of selected contours indicate that an asymmetric curved outline reduces the electrically visible lateral-mode features observed in more symmetric geometries, such as square resonators, while preserving high $k_t^2$ and $Q$. We further examine interconnection and device-area scaling, and benchmark the resulting performance against recently reported resonators for piezoelectric power conversion.

## Results

### Lateral-boundary apodization design concept

Figure 1 summarizes the device concept and implementation of lateral-boundary apodization. The resonator consists of a metallized 163° Y-cut $LiNbO_3$ plate mounted on a printed circuit board (PCB) and electrically connected through Au pads and wire bonds, as shown in Fig. 1A. The 163° Y-cut orientation enables strong thickness-shear (TS) excitation under the vertical electric field formed between the top and bottom aluminum electrodes. The cross-sectional schematic in Fig. 1E shows the $LiNbO_3$ plate, aluminum electrodes, Au contact pads, epoxy attachment layer, and PCB support, which together provide a mechanically stable platform for wire bonding and electrical characterization.

The square TS-mode plate exhibits a strong shear resonance but also a dense spurious spectrum near the main resonance and in the resonator inductive region towards the antiresonance, as shown in Fig. 1C, D. These spurious responses are attributed to lateral standing waves supported by the straight plate boundaries, which couple to the desired shear motion. In contrast, the shaped resonator in Fig. 1F–H retains the same material platform, electrode configuration, and PCB package, but uses an asymmetric curved outline to modify the lateral boundary conditions. This geometry modifies the lateral boundary conditions and is associated with reduced electrically visible spurious responses. Consequently, the shaped resonator shows reduced spurious-mode features in the admittance response, while retaining a dominant shear resonance and clearly defined $f_s$ and $f_p$ frequencies.

Overall, Fig. 1 establishes the central design strategy of this work: improving the spectral purity of a high-coupling $LiNbO_3$ TS resonator through lateral-boundary apodization. This approach uses blanket metal deposition followed by laser machining, avoiding fine electrode patterning or lithographic definition of the active resonator body while preserving strong TS-mode coupling. Detailed CAD geometries and boundary definitions of the tested resonators are provided in Fig. S1 and Tables S1–S3 in the Supplementary Information.

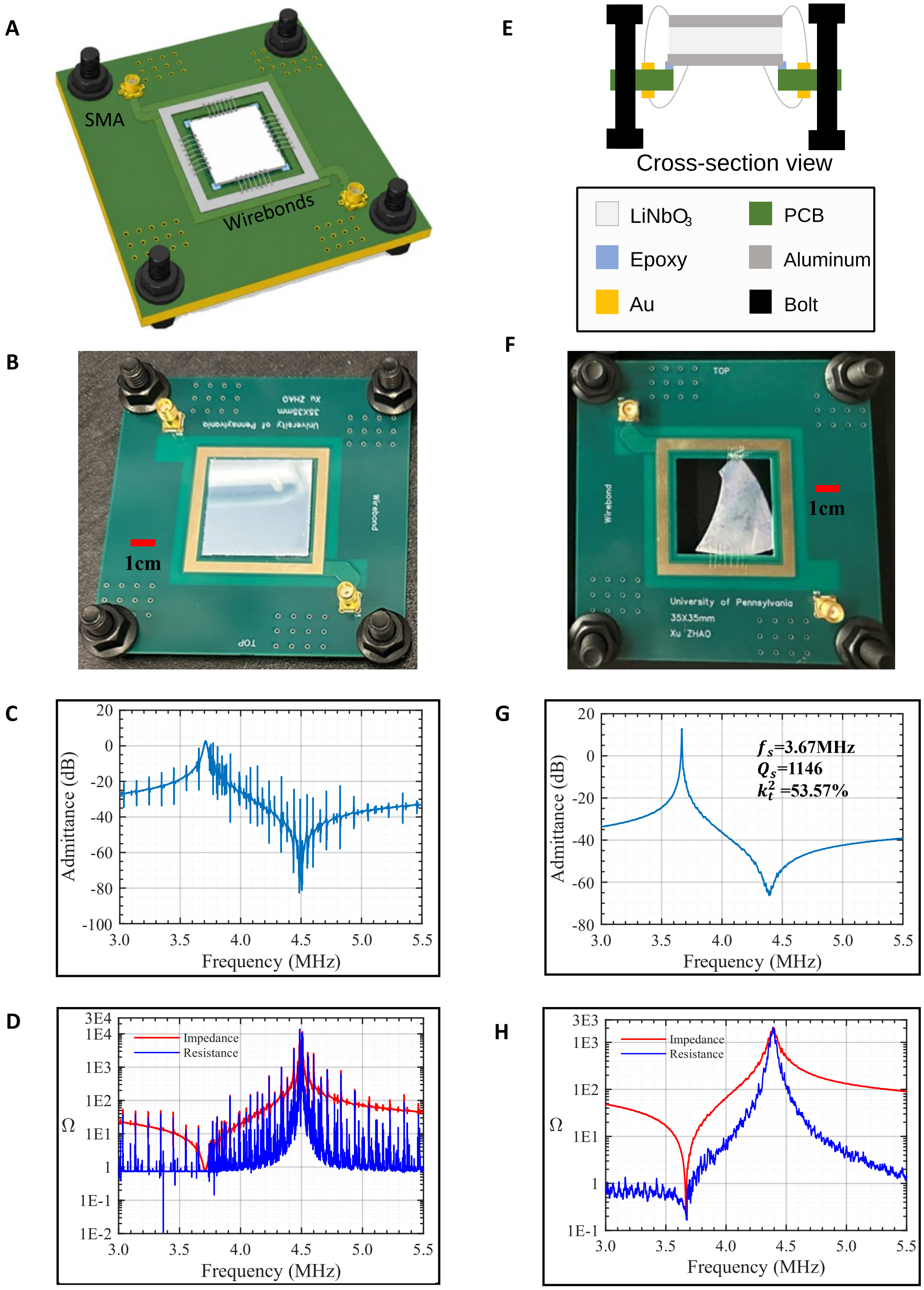

**Fig. 1. Design and realization of the apodized $LiNbO_3$ thickness-shear resonator.** **(A)** Three-dimensional schematic of the packaged resonator on the PCB. **(B)** Photograph of a square reference resonator. **(C)**

Measured admittance of the square resonator showing pronounced spurious responses near the thickness-shear resonance. **(D)** Measured impedance and resistance spectra of the square resonator, showing substantial fluctuations associated with spurious modes. **(E)** Cross-sectional view indicating the $LiNbO_3$ plate, aluminum electrodes, Au contact pads, epoxy, and PCB. **(F)** Photograph of a shaped resonator. **(G)** Measured admittance of the shaped resonator showing reduced spurious responses while retaining the main thickness-shear resonance. **(H)** Measured impedance and resistance spectra of the shaped resonator, showing a smoother response around the main resonance region.

## Finite-element analysis of square and shaped resonators

Finite-element simulations were performed in COMSOL Multiphysics to compare the vibration spectra of a square $LiNbO_3$ plate and a plate with the proposed asymmetric curved outline, as shown in Fig. 2A. The simulations used a frequency-domain piezoelectric finite-element model that couples solid mechanics and electrostatics. The $LiNbO_3$ domain was assigned anisotropic elastic, piezoelectric, and dielectric properties rotated to the 163° Y-cut coordinate system, and aluminum electrodes were included on the top and bottom surfaces to form the vertical excitation path. A harmonic voltage was applied between the two electrodes to excite the thickness-shear (TS) response, and the admittance was obtained from the simulated terminal current and applied voltage as ($Y = I/V$). To isolate the effect of the lateral resonator outline, the PCB, epoxy, and wire bonds were excluded from the model. Both resonators used the same electrode configuration, cross-sectional layer stack, and the same nominal area of 9 $cm^2$.

The simulated admittance spectra in Fig. 2C show a clear difference between the two geometries. In the square resonator, the main shear resonance is accompanied by a series of closely spaced spurious responses, labeled S1–S10. The electric-field-norm distributions in Fig. 2G reveal that these spurious modes exhibit nonuniform lateral standing-wave patterns with different spatial orders, confirming that they originate from parasitic lateral acoustic modes rather than the desired uniform TS response. These modes are supported by the straight and parallel boundaries of the square plate, which promote coherent lateral acoustic reflections and allow the parasitic modes to couple to the primary shear resonance.

In contrast, the shaped resonator retains the principal shear resonance while strongly suppressing the higher-order spurious responses. The asymmetric curved outline breaks the geometric symmetry of the plate and modifies the boundary conditions associated with lateral standing waves. As a result, the simulated admittance spectrum becomes substantially cleaner, with the main TS resonance preserved and the dense spurious features substantially reduced. These results provide two key design guidelines: the desired thickness-shear mode vibration should remain concentrated in the central active region to maintain strong overlap with the vertical electric field, and the lateral boundary should avoid long, parallel edges that can support high-$Q$ acoustic recirculation.

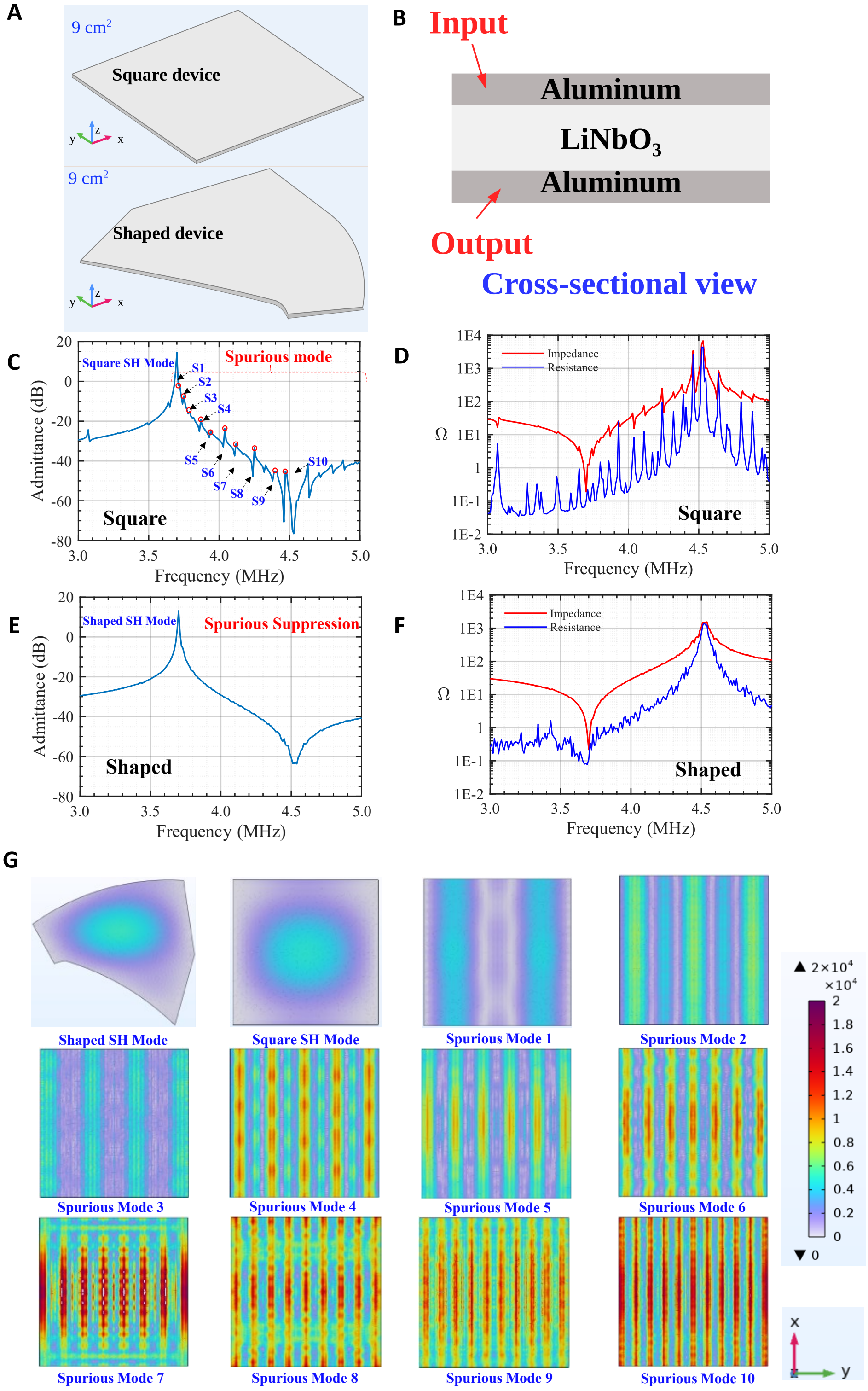

**Fig. 2. Simulated electrical responses and electric-field mode shapes of square and apodized resonators.** **(A)** Square and shaped $LiNbO_3$ resonator geometries with the same nominal area of 9 $cm^2$. **(B)** Cross-sectional schematic showing the $LiNbO_3$ plate between the top and bottom aluminum electrodes used for electrical excitation. **(C)** Simulated admittance spectrum of the square resonator, showing the principal thickness-shear mode together with a dense sequence of spurious modes, S1–S10. **(D)** Simulated impedance and resistance spectra of the square resonator. **(E)** Simulated admittance spectrum of the shaped resonator, showing preservation of the principal thickness-shear resonance with substantially reduced spurious features. **(F)** Simulated impedance and resistance spectra of the shaped resonator. **(G)** Electric-field-norm distributions of the principal thickness-shear modes of the shaped and square resonators and representative spurious modes of the square resonator.

## Experimental comparison of selected resonator shapes

To examine selected lateral-boundary contours experimentally, resonators with different lateral outlines were fabricated and characterized, as shown in Fig. 3. Because the devices in Fig. 3 used different packaged interconnections from the device in Fig. 1F, their absolute admittance levels should not be directly compared; this wire-bonding effect is discussed further in Fig. 4. The corresponding CAD definitions and fabrication/interconnection conditions are summarized in Fig. S1 and Tables S1–S3. The square reference device exhibits a strong thickness-shear resonance at approximately 3.71 MHz with a high $k_t^2$ of 56.94%. However, its admittance spectrum contains numerous spurious peaks and notches between 3.6 and 5.0 MHz, consistent with the lateral parasitic modes predicted by finite-element simulation.

In contrast, the shaped resonators show substantially cleaner admittance responses while maintaining high $k_t^2$. Across the measured shaped devices, the $f_s$ remains clustered near 3.65–3.67 MHz, and the extracted $k_t^2$ remains high, ranging from 53.60% to 58.50%. These results indicate that the primary TS resonance is mainly determined by the plate thickness and shear acoustic velocity, while the lateral outline influences electrically visible spurious responses. The circular-sector-like and asymmetric designs generally show reduced lateral-mode features relative to the square plate, although the degree of reduction varies among geometries.

The small variations in resonance frequency and electromechanical coupling among the shaped devices are attributed to differences in active area, boundary loading, and the overlap between the electrode field and the shear-displacement distribution. Among the tested geometries, Shape 5 exhibits the lowest electrically visible lateral-mode structure while maintaining a high coupling coefficient of ($k_t^2$ = 53.60%). This improvement is likely due to its strongly asymmetric and tapered lateral outline, which may alter lateral acoustic reflections and reduce the electrical visibility of parasitic modes. In contrast, the square resonator provides parallel boundaries that support strong lateral resonances, resulting in pronounced spurious responses. Therefore, the results suggest that the selected asymmetric contours can be a useful approach for reducing electrically visible spurious responses without significantly degrading the large intrinsic coupling of the $LiNbO_3$ TS mode.

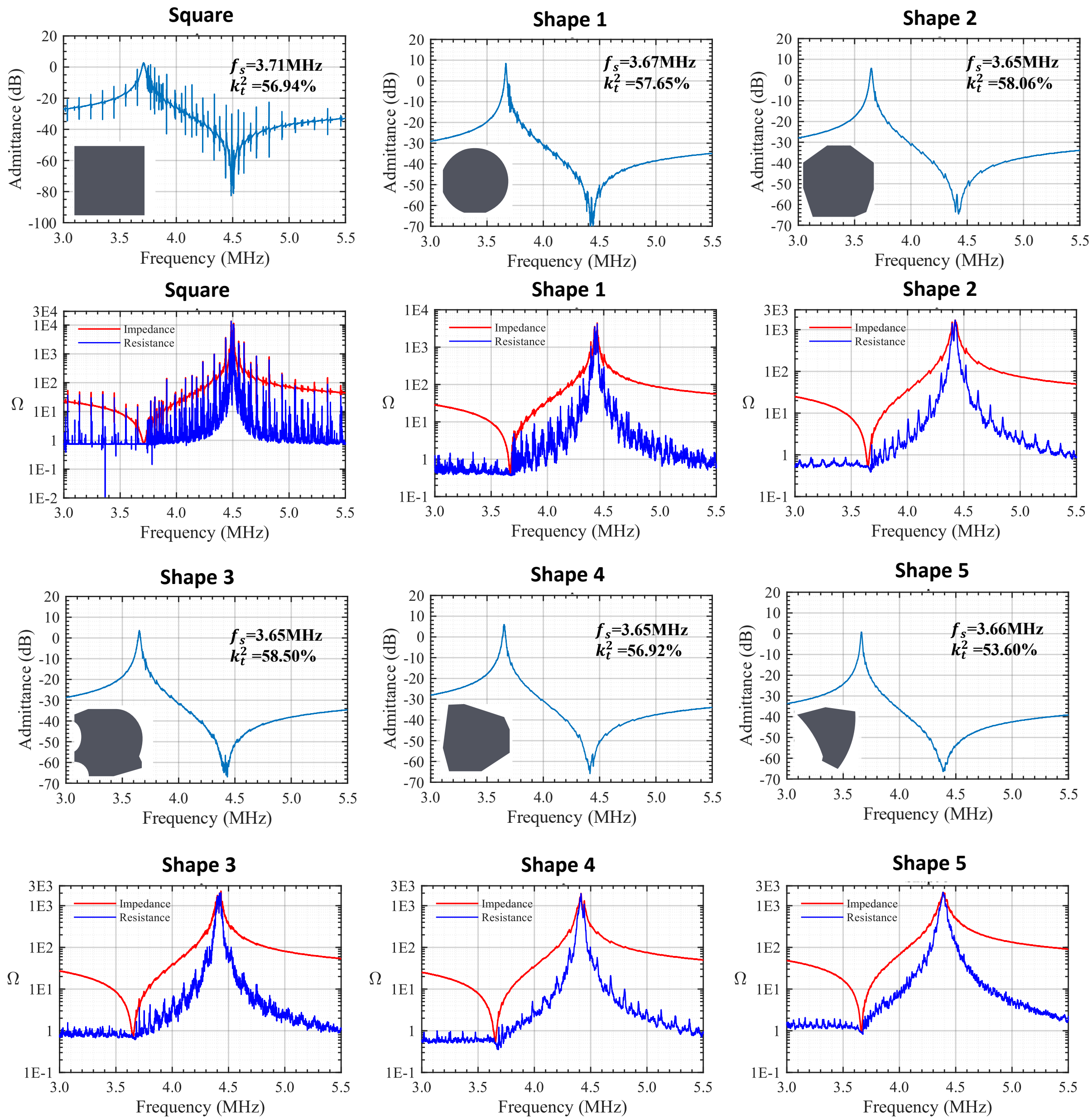

**Fig. 3. Measured electrical responses of the square reference and five apodized $LiNbO_3$ thickness-shear resonators.** For each resonator, the upper panel shows the measured admittance spectrum with the extracted $f_s$, and $k_t^2$, while the lower panel shows the corresponding impedance and resistance spectra. The square reference exhibits pronounced spurious responses near and above the main thickness-shear resonance, whereas the shaped resonators show substantially reduced spurious features while maintaining high $k_t^2$. Shape 5 has the cleanest resistance spectrum near $f_s$. Insets show the corresponding resonator outlines. Photographs of the packaged resonators with different geometries are shown in Fig. S4.

### Influence of interconnects

Figure 4 examines the effect of wire-bond configuration on the measured response of shaped $LiNbO_3$ TS resonators. Figures 4A and 4B compare 163° Y-cut shaped devices with different wire-bond configurations. With fewer wire bonds, the device exhibits a clean resonance near 3.66 MHz but a relatively low $Q_s$ of 271. After increasing the number of wire bonds and improving the electrical connection to the electrode, the $f_s$ remains nearly unchanged at 3.67 MHz, while the $Q_s$ increases to 1146. This improvement is consistent with reduced contact resistance and more uniform current distribution across the electrode. Additional photographs of the wire-bonded devices are provided in Fig. S5, with the corresponding bond counts and attachment conditions summarized in Table S3. These results indicate that, alongside lateral-boundary apodization, interconnection design is also important for preserving the intrinsic low-loss acoustic response. High spectral purity can be maintained across different interconnection configurations, while optimized wire bonding is critical for achieving high $Q$ in the packaged device.

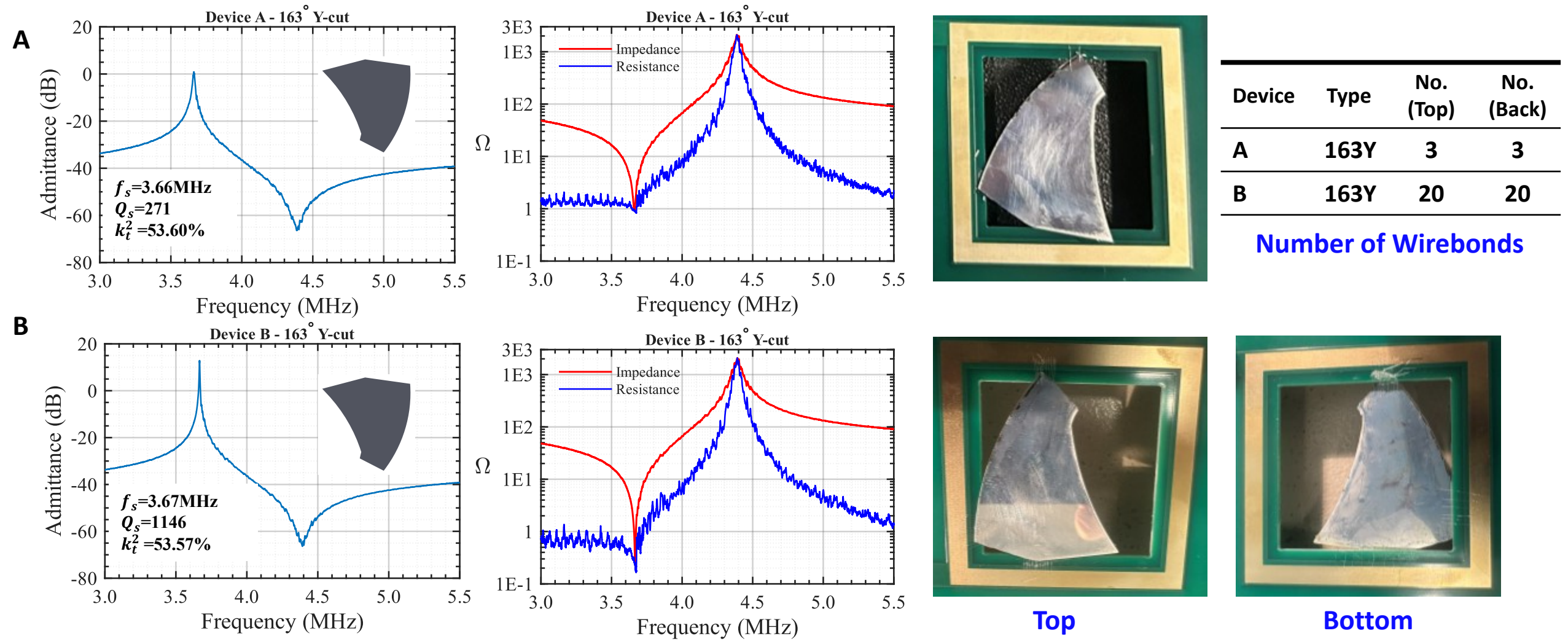

**Fig. 4. Effect of wire-bond configuration on the measured response of apodized $LiNbO_3$ thickness-shear resonators.** **(A)** Device A with three wire bonds on each electrode side, showing the measured admittance, impedance, and resistance spectra together with a photograph of the packaged resonator. **(B)** Device B with twenty wire bonds on each electrode side, showing the corresponding admittance, impedance, and resistance spectra and top- and bottom-side photographs of the resonator. Increasing the number of wire bonds substantially increases the extracted series quality factor, (**$Q_s$**), from 271 to 1146 while leaving the series resonance frequency, (**$f_s$**), nearly unchanged.

**Area scaling of shaped resonators**

Figure 5 examines how resonator area influences the electromechanical response of apodized TS-mode $LiNbO_3$ resonators. Devices with the same general asymmetric outline but different active areas of 40, 100, 190, 290, and 560 $mm^2$ were fabricated and measured, as shown in Fig. 5A, B. Additional topside and backside photographs of representative shaped resonators are provided in Fig. S6, showing the different lateral outlines used for the spurious-mode suppression and area-scaling studies. All area-scaled devices retain the same normalized Shape-5 boundary, with the uniform scaling rule and corresponding dimensions summarized in Table S4. All devices exhibit a dominant TS $f_s$ near 3.67–3.69 MHz and a clearly defined antiresonance near 4.2–4.4 MHz. The small variation in resonance frequency confirms that the main TS resonance is primarily determined by the plate thickness and shear acoustic velocity, rather than by the lateral device area.

As the resonator area increases, the admittance peak becomes sharper, and the extracted $Q$ improves, as summarized in Fig. 5C. The series quality factor increases from 90 for the 40 $mm^2$ device to 1146 for the 560 $mm^2$ device. This improvement is especially important for piezoelectric DC-DC power-converter operation, because the converter is expected to operate near the $f_s$, where the motional impedance is minimized and $Q_s$ directly reflects the acoustic energy storage relative to dissipation. The strong area dependence of the series quality factor is consistent with an increased relative contribution of edge- and boundary-related loss in smaller devices. As the device area increases, a larger fraction of the acoustic energy is stored in the central active region, reducing the relative contribution of edge- and boundary-perturbed regions.

The $k_t^2$ also remains high across the shaped devices and increases with area, reaching 53.57% for the 560 $mm^2$ resonator. This indicates that area scaling improves $Q$ without compromising the strong $k_t^2$ of the TS mode. The largest device therefore provides the highest series quality factor and series-resonance FoM within the measured area series, with ($f_s$ = 3.67) MHz, ($Q_s$ = 1146), and ($k_t^2$ = 53.57%), corresponding to a series-resonance figure of merit ($k_t^2$ $Q_s$ = 614). The concurrent increase in the extracted $k_t^2$and $Q$ is also consistent with prior BAW resonator modeling, showing that both $k_t^2$ and $Q$ can depend on the area-to-perimeter ratio (A/p), where stronger edge clamping in small-A/p resonators reduces the effective electromechanical response[23]. In our shaped TS-mode $LiNbO_3$ devices, increasing the resonator area reduces the relative contribution of edge- and boundary-perturbed regions, thereby improving the participation of the central TS mode and increasing the measured $k_t^2$.

Importantly, the improvement in $Q$ is achieved while maintaining reduced lateral-mode features in the measured admittance response. Even for the largest 560 mm² resonator, the dense spurious responses observed in the square reference device do not reappear. These results show that reduced electrically visible spurious features are retained across the tested areas and that larger-area resonators can simultaneously provide high $Q$, high coupling, and improved spectral purity. The corresponding area-dependent $R$, uninterrupted inductive bandwidth, and effective-$L$ ranges are summarized in Table S5.

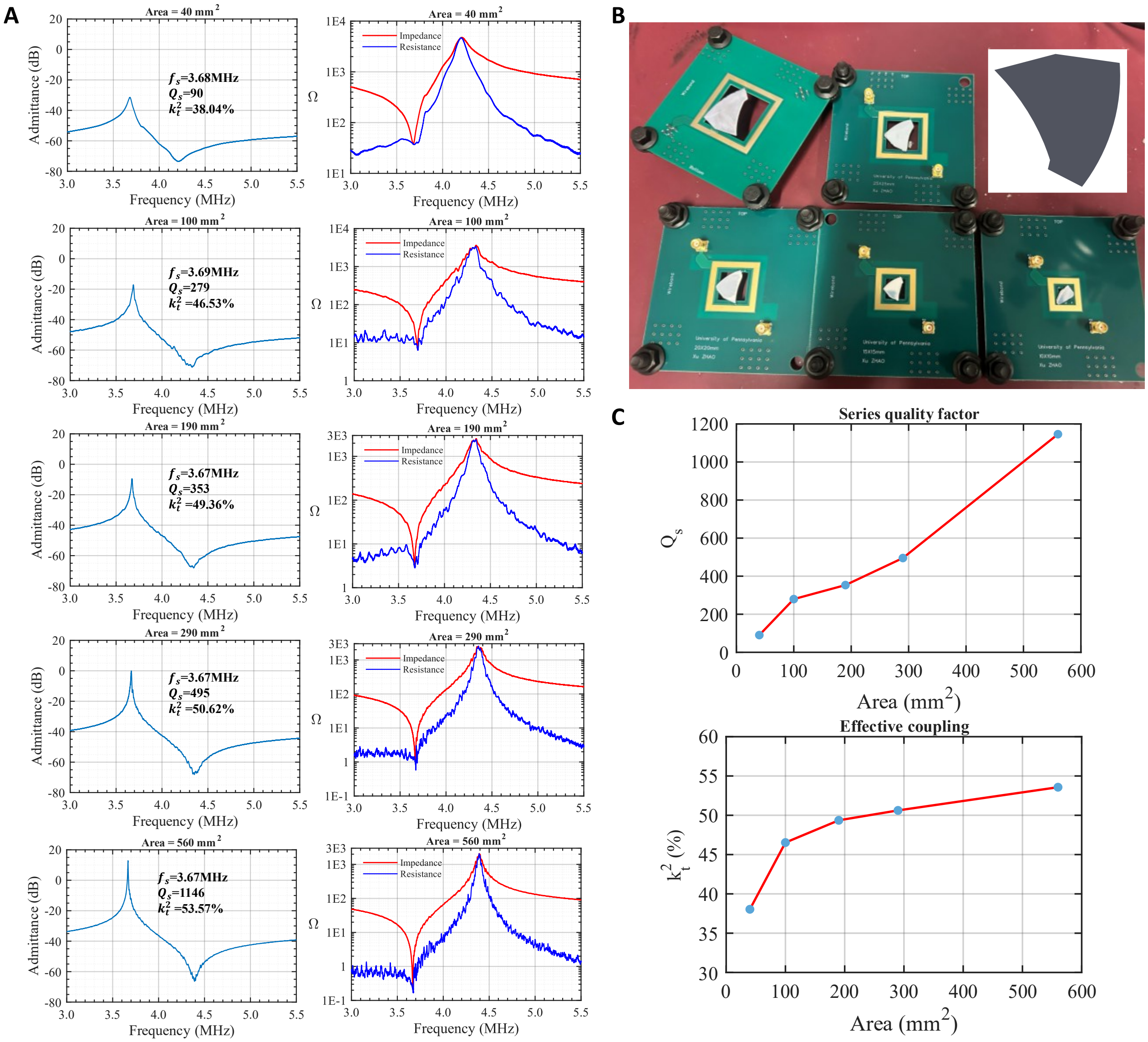

**Fig. 5. Area dependence of apodized $LiNbO_3$ thickness-shear resonators. (A)** Measured admittance, impedance, and resistance spectra for shaped resonators with areas of 40, 100, 190, 290, and 560 mm$^2$. The extracted series resonance frequency, ($f_s$), series quality factor, ($Q_s$), and effective electromechanical coupling coefficient, ($k_t^2$ ), are indicated for each device. **(B)** Photograph of packaged resonators with different device areas; the inset shows the common asymmetric device outline used for the area-scaling study. **(C)** Extracted ($Q_s$) and ($k_t^2$ ) as functions of resonator area. Larger devices exhibit higher ($Q_s$) while maintaining high effective electromechanical coupling and reduced lateral-mode features.

## Equivalent-circuit fitting and Bode-Q extraction

A representative high-performance shaped resonator was fitted with a modified Butterworth-van Dyke (MBVD) equivalent circuit to validate the extracted resonator parameters[24]. The measured admittance magnitude and phase agree closely with the model over the 2.0-6.0 MHz range, including the resonance, antiresonance and phase transition. The fitted motional parameters are $L_m$= 8.9 μH, $C_m$ = 211 pF and $R_m$ = 0.155 Ω, with $C_0$ = 486 pF, $R_s$ = 0.025 Ω and $R_0$ = 1.97 Ω. These values are consistent with a strongly coupled, low-resistance acoustic resonance suitable for power-converter operation.

The same device was analyzed using impedance, resistance and Bode-$Q$ extraction. The extracted series and parallel resonances were $f_s$ = 3.667 MHz and $f_p$ = 4.392 MHz, with $k_t^2$ = 53.57%. The Bode quality factor extracted at the series resonance ($f_s$ = 3.667 MHz) was ($Q_{B(f_s)}$=1621), whereas the maximum Bode quality factor was 2896 at 3.675 MHz. The difference between 3-dB $Q$ and Bode-$Q$ is expected because the two metrics weight loss and phase slope differently. For benchmarking converter resonators, both values are informative. The MBVD fit, direct admittance extraction, and Bode-$Q$ analysis collectively confirm the high coupling and low-loss operation of the shaped TS-mode resonator.

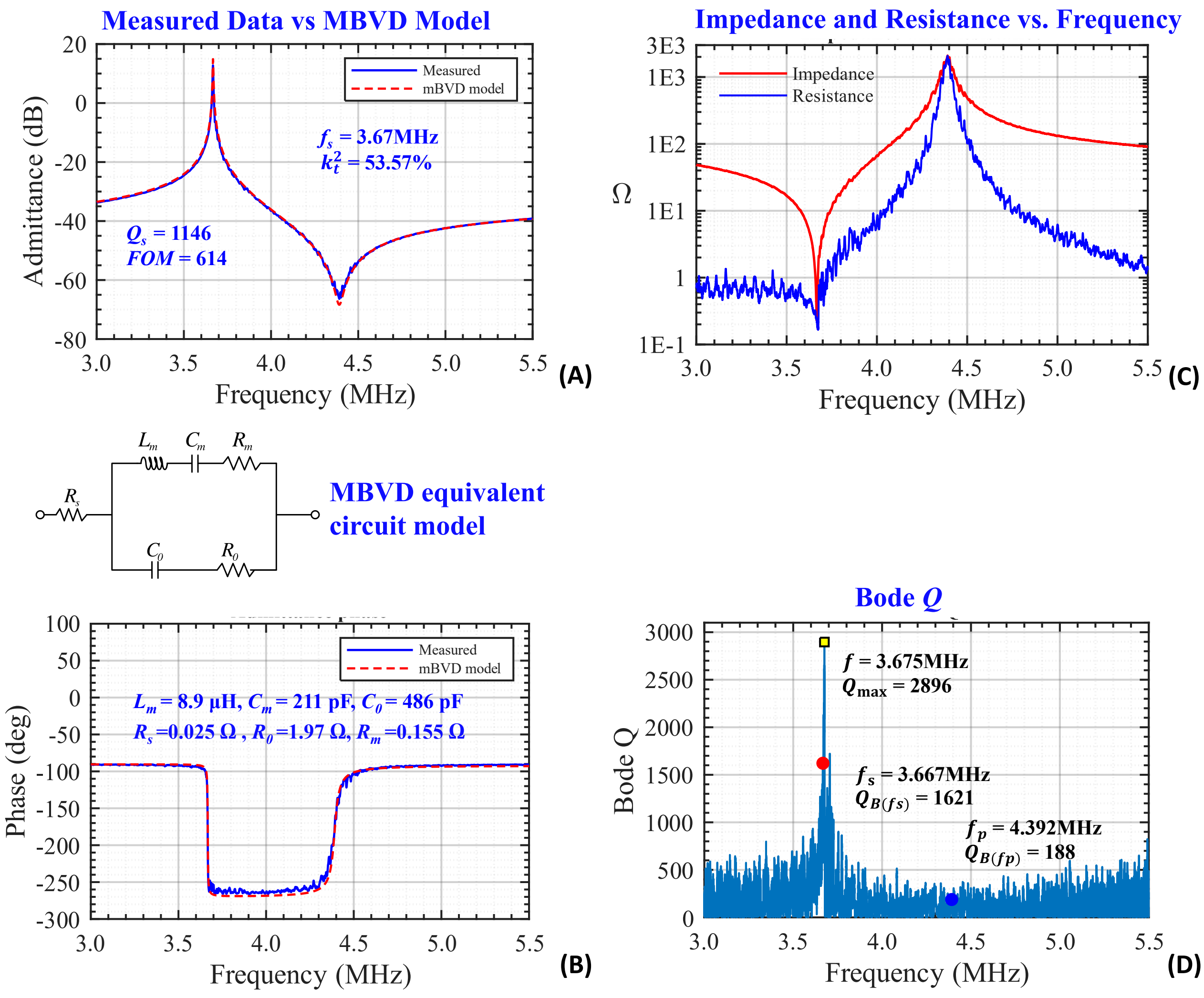

**Fig. 6. Measured response, MBVD fitting, and Bode-*Q* analysis of a representative apodized 163° Y-cut $LiNbO_3$ thickness-shear resonator.** **(A)** Measured and MBVD-fitted admittance magnitude of the resonator, with the extracted $f_s$, effective electromechanical coupling coefficient, ($k_t^2$), $Q_s$, and figure of merit, ($k_t^2 Q_s$). **(B)** Measured and MBVD-fitted admittance phase, together with the modified Butterworth–Van Dyke (MBVD) equivalent-circuit model and fitted circuit parameters. **(C)** Measured impedance and resistance spectra, showing the series and parallel resonances and the inductive region between $f_s$ and $f_p$. **(D)** Extracted Bode-*Q* as a function of frequency, showing a maximum Bode *Q* of 2896 near 3.675 MHz. The corresponding device exhibits ($k_t^2$ = 53.57%) and ($Q_s$ = 1146) by 3-dB extraction.

## Performance comparison

Table 1 compares the present apodized TS-mode resonator with representative resonators reported for piezoelectric DC-DC power conversion[7,16,17,19,20,25]. The present device achieves the highest effective electromechanical coupling among the listed resonators while maintaining a high series quality factor and a low resistance-residual metric. For the representative 560 mm² shaped resonator shown in Fig. 6, the extracted series quality factor is ($Q_s$ = 1146), giving a series-resonance figure of merit of ($k_t^2$ $Q_s$ = 614). The same device exhibits a Bode quality factor of 2896, corresponding to a Bode-$Q$ based figure of merit of ($k_t^2$ $Q_B$ =1552).

To quantitatively compare the spurious response, the residual energy per MHz was extracted from the resistance spectrum, as described in detail in Figs. S17-S19 in the Supplementary Information. This metric measures the integrated squared residual between the measured curve and the smooth main mode fitting curve, normalized by the analyzed frequency span. Lower values indicate smaller resistance fluctuations relative to the smooth main-mode response. Because the analysis contains user-defined valley/peak search windows and an empirically constructed baseline, residual per MHz is treated as a comparative continuous metric rather than an absolute definition of a spurious-free resonator.

The present device shows a residual energy per MHz of 0.014, among the lowest values in the benchmarked resistance spectra. The relative result remains robust under variations of the PCHIP baseline parameters, as detailed in the Supplementary Information. As an independent electrical cross-check, the representative 560 mm² Shape-5 device also maintains an uninterrupted inductive interval of approximately 723 kHz. Together with the high effective coupling and quality factor, these results support reduced electrically visible lateral-mode features in the shaped TS-mode $LiNbO_3$ resonator.

**Table 1. Comparison of representative resonators for piezoelectric DC-DC power conversion**

| Reference | Mode | $f_s$ (MHz) | $k_t^2$ | $Q_s$ | FoM ($Q_s$) | Bode $Q$ /$Q_m$ | FoM (Bode $Q$/$Q_m$) | Residual per MHz |
|---|---|---|---|---|---|---|---|---|
| This Work | TS-LN | 3.67 | 53.6% | 1146 | 614 | 2896 | 1552 | 0.014 |
| Braun et al.[7] | TE-LN | 6.8 | 29% | NA | NA | NA | NA | 0.206 |
| Nguyen et al.[16] | TS-LN | 5.9 | 45% | NA | NA | 3500 | 1575 | 0.094 |
| Nguyen et al.[17] | TE-LN | 10.1 | 30% | NA | NA | 4000 | 1200 | 0.016 |
| Yao et al.[19] | TE-LT | 6.25 | 8.8% | NA | NA | 1698 | 149 | 0.032 |
| Yao et al.[20] | TE-LN | 19.23 | 29% | 1282 | 372 | 3187 | 928 | 0.176 |
| Zhao et al.[25] | TS-LN | 3.6 | 53.5% | 711 | 381 | 2629 | 1407 | 0.012 |
| Yao et al.[26] | RE-LN | 0.313 | 16.8% | NA | NA | 2500$^m$ | 420$^m$ | NA |
| Daniel et al.[12] | RE-PZT | 0.174 | 42.5% | NA | NA | 1198$^m$ | 509$^m$ | NA |
| Boles et al.[14] | TE-PZT | 0.61 | 31% | NA | NA | 2500$^m$ | 775$^m$ | NA |
| Yao et al.[27] | TE-AlN | 13.52 | 6.1% | NA | NA | 1677 | 103 | NA |

Where TE = thickness extension; TS = thickness shear; RE = radial extension; LN = lithium niobate; LT = lithium tantalate; PZT = lead zirconate titanate; AlN = aluminum nitride and NA = not available. $Q_s$ denotes the series-resonance quality factor, whereas $Q_m$ denotes the mechanical/motional quality factor as reported in the corresponding reference. Superscript "m" indicates values reported as $Q_m$. Residual per MHz is a continuous comparative smoothness metric obtained from the valley-to-peak log-resistance residual relative to the monotonic PCHIP baseline; no universal categorical threshold is assigned. Further methodological details and robustness tests are provided in the Supplementary Information.

## Discussion

The results support lateral-boundary apodization as an approach for reducing the electrically visible response of spurious modes in high-coupling TS-mode $LiNbO_3$ resonators. Unlike multi-finger excitation, this approach leaves the active vertical electric field largely intact, thereby preserving the strong coupling of the shear mode. The strategy is also compatible with a lithography-free fabrication flow because the resonator outline can be defined by laser cutting after blanket electrode deposition. This feature is useful for rapid design iteration and for large-area resonators intended for DC-DC power conversion, where fine lithographic patterning is costly and not always necessary.

One possible physical interpretation is the modification of lateral standing-wave boundary conditions. In square plates, long parallel edges allow many lateral modes to satisfy resonance conditions near the principal TS resonance. These modes are coupled to the electrical port and appear as spurious peaks and notches in admittance. Curved and asymmetric boundaries alter lateral acoustic reflections, thereby reducing the electrical visibility of unwanted modes. The finite-element mode shapes in Fig. 2 and the measured spectra in Fig. 3 are consistent with this interpretation. The same-area simulation isolates boundary geometry within the model; the fabricated contour, scaling, interconnection, and repeatability studies address complementary questions rather than constituting a controlled experimental geometric sensitivity study. An independently fabricated Shape-5 resonator reproduces the same overall electrical response, as shown in Figs. S13–S14 and Table S8.

Packaging is a second critical factor. The strong improvement in $Q$ with increased wire bonding indicates that electrical access and boundary loading can limit the measured resonator performance. The interconnect must provide low series resistance while avoiding stiff or nonuniform mechanical constraints that can seed spurious modes or lead to anchor loss. The shaped devices with improved wire bonds achieve substantially higher $Q$ than sparsely bonded devices, showing that intrinsic acoustic designs and package should be treated as a coupled system. The PCB-fixture contribution is further quantified by thru de-embedding in Figs. S10–S12 and Table S7.

Area scaling further clarifies the trade-off between motional impedance and boundary-related dissipation. Larger devices store more acoustic energy in the interior relative to the perimeter and therefore tend to exhibit higher $Q$. At the same time, a large area can increase capacitance and alter impedance matching in a converter. The 560 $mm^2$ device provides the highest measured series quality factor and FOM within this area series, but future converter-level optimization should choose the resonator area together with the target voltage conversion ratio, load range, and operating frequency.

This work focuses on small-signal resonator performance rather than a complete converter demonstration. Excitation-level sensitivity from −15 to +20 dBm is additionally evaluated in Figs. S15–S16 and Table S9, showing only small resonance-frequency variation over the measured range. High-power operation may introduce nonlinear damping, self-heating, voltage breakdown or contact degradation[12,28,29], all of which must be evaluated before system deployment. In addition, temperature variations may shift the resonance frequency and affect resonator stability[30–32], and future work would examine thermal behavior and temperature-compensation strategies for converter operation. Nevertheless, the combination of $k_t^2$ above 53%, $Q_s$ approaching 1100, reduced electrically visible lateral-mode features, and an MBVD model that accurately captures the measured response provides a strong foundation for future high-power converter integration.

## Materials and methods

### Device design and finite-element simulation

The resonators were designed as metalized $LiNbO_3$ plates operating in a TS mode near 3.65 MHz. The primary comparison used a square plate and an asymmetric curved plate with the same nominal area of 9 $cm^2$. Frequency-domain finite-element simulations were conducted in COMSOL Multiphysics by coupling the Solid Mechanics and Electrostatics physics interfaces through the Piezoelectricity Multiphysics coupling. The $LiNbO_3$ plate was modeled as an anisotropic piezoelectric solid with material tensors rotated to the target crystal cut, and the aluminum electrodes were included as conductive layers defining the electric field direction across the plate thickness. A small-signal harmonic voltage of 1 V was applied to the top electrode, while the bottom electrode was grounded. The complex admittance was then calculated from the simulated terminal current as (Y = I/V). Mechanically free boundary conditions were applied to the exposed plate surfaces and lateral edges to isolate the intrinsic resonator response. Electric field norm and displacement mode shapes were extracted at the resonance peaks to distinguish the principal TS mode from lateral standing wave spurious modes.

### Lithography-free fabrication

As shown in Fig. S2 in the Supplementary Information, 0.5 mm thick 163° Y-cut $LiNbO_3$ wafers were cleaned before metal deposition. Aluminum electrodes were deposited (1.0 μm) on the top and bottom surfaces to form a vertical electric-field excitation path. The resonator outlines were subsequently defined by UV excimer laser micromachining using an IPG IX-255 Excimer Laser Micromachining system (LMM-01, Singh Center for Nanotechnology) equipped with a 193 nm UV excimer laser. For through-cutting, the metallized $LiNbO_3$ samples were mounted on a sacrificial carrier wafer or aluminum sheet and secured during processing. CAD-defined resonator outlines were cut using a nominal target energy of 7.0 mJ, a beam size of 10 × 10 $\mu m^2$, a repetition rate of 800 Hz, a pulse spacing of 1.0 μm, and 30 passes per shape. This laser-cutting process enabled square and shaped plates to be realized without photolithographic patterning of the active resonator body.

### Electrical measurement

The resonators were characterized using a vector network analyzer (Keysight, P9374A) in a 50 Ω environment. Devices were mounted on PCBs with SMA connectors and measured in a two-port configuration as shown in Fig. S3 in the Supplementary Information. A short-open-load-through calibration was performed to the ends of the cables before measurement to reduce the influence of cables, thereby improving the accuracy of the extracted electrical response. The PCB-based connectors and electrical routing are not de-embedded and are also included in the overall measured responses. The measured S-parameters were converted to admittance to identify the series and parallel resonance frequencies. Measurements were performed over frequency windows that included the main TS resonance and the surrounding spurious-mode region. Post-calibration thru verification is provided in Fig. S7.

### Parameter extraction

The series resonance frequency $f_s$ was taken at the admittance maximum, and the parallel resonance frequency $f_p$ was taken at the admittance minimum. The resonator-limited efficiency expression in Eq. (1) follows prior analyses of piezoelectric power conversion[7]. The effective electromechanical coupling coefficient was calculated from $f_s$ and $f_p$ using Eq. (2), following the convention commonly used for strongly coupled acoustic resonators. The series quality factor was extracted from the 3-dB bandwidth around $f_s$ using Eq. (3). The Bode quality factor was calculated from the phase slope of the measured reflection coefficient using Eq. (4), following the formulation of Feld et al.[33]. A modified Butterworth-van Dyke model containing a motional branch and static capacitance branch was fitted to the measured admittance magnitude and phase data.

$$k_t^2 = \frac{\pi^2}{8}\frac{{f_p}^2 - {f_s}^2}{{f_s}^2} \tag{2}$$

$$Q_s = \frac{f_s}{\Delta f_{s,3dB}} \tag{3}$$

$$Q_B = \omega\frac{d\phi s_{11}}{d\omega}\frac{|S_{11}|}{1 - |S_{11}|^2} \tag{4}$$

### Data availability

The data supporting the findings of this study are available within the article and its supplementary information. Additional raw test data is available from the corresponding author upon reasonable request.

## Acknowledgements

The fabrication and characterization of the resonators were supported by facilities and infrastructure at the University of Pennsylvania. The authors acknowledge the Singh Center for Nanotechnology, which is supported by the NSF National Nanotechnology Coordinated Infrastructure Program (NNCI-1542153).

## Author contributions

X.Z. and R.H.O. conceived the device concept. X.Z. designed, fabricated, simulated, and characterized the resonators. X.Z. and S.Y. quantitatively compared the spurious response. X.Z. drafted the manuscript. S.Y. and R.H.O. revised the manuscript, and R.H.O. supervised the research. All authors discussed the results and approved the final manuscript.

## Conflict of interest

The authors declare no conflict of interest.

## Supplementary information

The online version contains supplementary material available in Supplementary Information.

# Supplementary Information for

# Lithography-free, apodized thickness-shear $LiNbO_3$ resonators for piezoelectric DC-DC power conversion

Xu Zhao[1], Shun Yao[1] and Roy H. Olsson III[1]*

[1]Department of Electrical and Systems Engineering, University of Pennsylvania, Philadelphia, PA, USA

*Corresponding author Email: rolsson@engineering.upenn.edu

## Lateral-boundary apodization and contour definitions

### Comparison of selected lateral contours

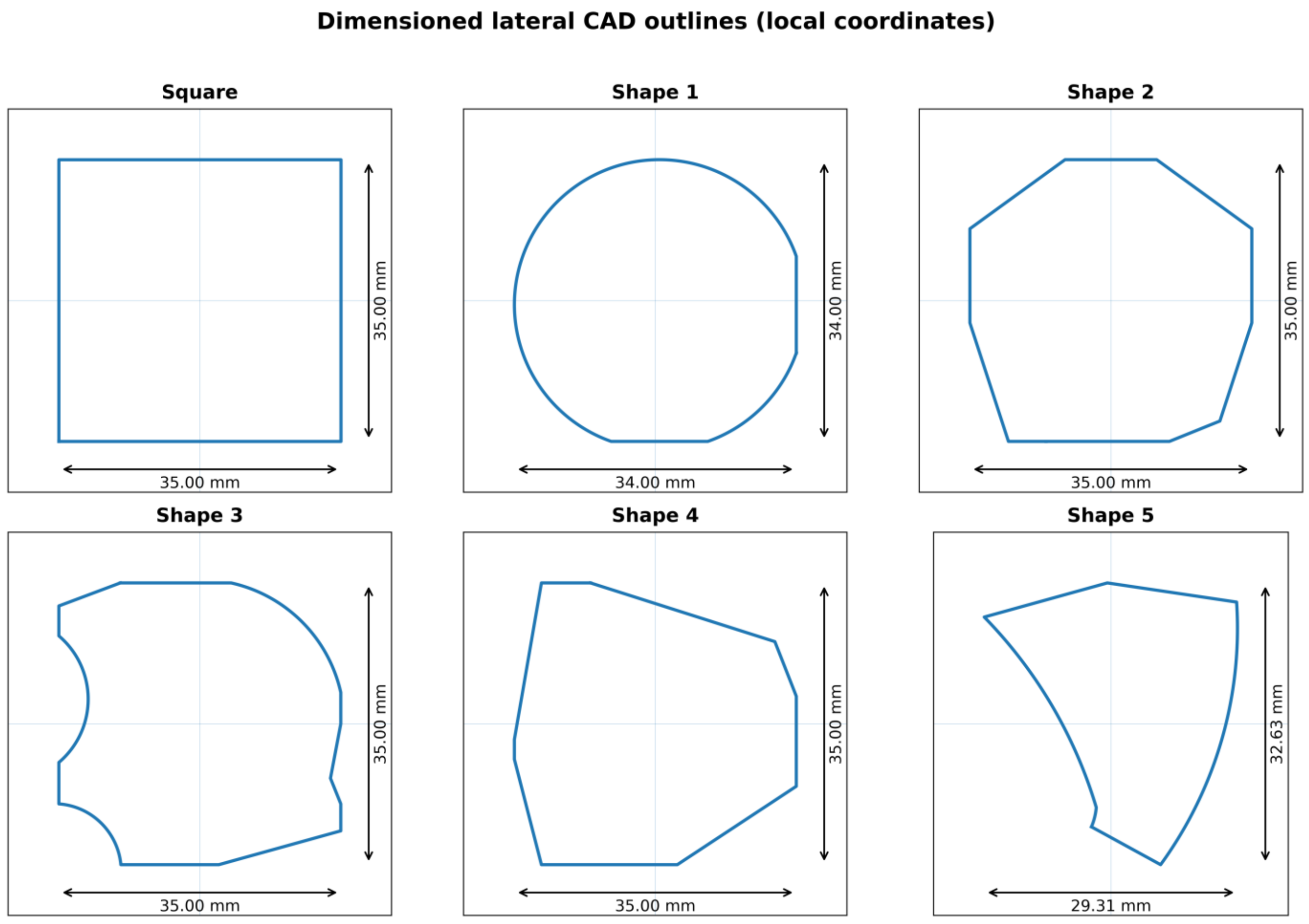

***Figure S1.*** *Dimensioned CAD outlines used in the shape-comparison study. Overall, lateral width and height are shown for each device. Detailed boundary coordinates, straight-edge angles, and arc radii are provided in Tables S1–S2.*

**Design definition:** In this work, "lateral-boundary apodization" refers to modification of the physical $LiNbO_3$ plate contour to reduce electrically visible spurious responses, as distinct from electrode apodization or a formal multi-parameter geometric optimization. The tested outlines were selected to explore variations in parallel-edge length, curvature, taper, and asymmetry while retaining the thickness-shear active region. The square reference provides the highly symmetric baseline. Shape 1 removes most straight parallel edges but remains approximately circular and therefore highly symmetric, whereas Shapes 2–5 introduce progressively more irregular, tapered, concave, and/or curved boundaries.

**CAD source and units:** All dimensions below were extracted directly from the supplied DXF files. The DXF insertion units are millimeters. The numerical boundary tables below use a local coordinate system centered at the bounding box of each individual outline. The full vector geometry is directly reproducible.

**Geometric summary**

| Device | Width (mm) | Height (mm) | Area (mm²) | Perimeter (mm) | Area/Perimeter | Boundary |
|---|---|---|---|---|---|---|
| Square | 35.00 | 35.00 | 1225.00 | 140.00 | 8.75 | 4 lines |
| Shape 1 | 34.00 | 34.00 | 946.47 | 109.50 | 8.64 | 2 lines + 2 arcs |
| Shape 2 | 35.00 | 35.00 | 1046.22 | 119.01 | 8.79 | 10 lines |
| Shape 3 | 35.00 | 35.00 | 1036.49 | 126.12 | 8.22 | 10 lines + 3 arcs |
| Shape 4 | 35.00 | 35.00 | 986.52 | 118.86 | 8.30 | 10 lines |
| Shape 5 | 29.31 | 32.63 | 563.39 | 99.49 | 5.66 | 3 lines + 3 arcs |

***Table S1.*** *Summary of the lateral CAD geometries. *Straight-edge orientation is reported relative to the local +x direction and reduced to 0–180°. These angles provide the explicit taper/edge-direction information needed to reproduce the polygonal portions of each outline.*

**Geometric design variables:** The CAD comparison spans overall width/height, active plate area, perimeter, straight-edge orientation, taper, concavity, and boundary curvature. These quantities are reported as geometric descriptors rather than as independently optimized variables. The present data provide a qualitative boundary-selection rationale—reducing parallel reflection paths and introducing asymmetry—rather than a unique optimum set of radii or taper angles. A systematic one-variable-at-a-time sensitivity study or numerical objective-function optimization was not performed in the present work and should not be implied. A practical workflow is to select plate thickness and area, compare candidate contours under matched numerical conditions, and assess the fabricated device and its interconnections.

**Boundary coordinates and segment definitions**

The following table lists the closed boundary in sequential order. Coordinates are in millimeters in the local coordinate system defined above. For straight segments, the segment length and direction are reported. For curved segments, the radius, sweep angle, and arc-center coordinates are reported. This representation provides a reproducible geometric definition of each device.

| Device | Seg. | Type | Start (x, y) mm | End (x, y) mm | Length / angle or arc definition |
|---|---|---|---|---|---|
| Square | 1 | Line | (-17.500, -17.500) | (17.500, -17.500) | L=35.000 mm; θ=0.00° |
| Square | 2 | Line | (17.500, -17.500) | (17.500, 17.500) | L=35.000 mm; θ=90.00° |
| Square | 3 | Line | (17.500, 17.500) | (-17.500, 17.500) | L=35.000 mm; θ=0.00° |
| Square | 4 | Line | (-17.500, 17.500) | (-17.500, -17.500) | L=35.000 mm; θ=90.00° |
| Shape 1 | 1 | Line | (17.000, -6.331) | (17.000, 5.331) | L=11.662 mm; θ=90.00° |
| Shape 1 | 2 | Arc | (17.000, 5.331) | (-5.331, -17.000) | R=17.500 mm; sweep=231.07°; center= (0.500, -0.500) |

| Device | Seg. | Type | Start (x, y) mm | End (x, y) mm | Length / angle or arc definition |
|---|---|---|---|---|---|
| Shape 1 | 3 | Line | (-5.331, -17.000) | (6.331, -17.000) | L=11.662 mm; θ=0.00° |
| Shape 1 | 4 | Arc | (6.331, -17.000) | (17.000, -6.331) | R=17.500 mm; sweep=51.07°; center= (0.500, -0.500) |
| Shape 2 | 1 | Line | (-8.049, -17.500) | (-12.714, -17.500) | L=4.666 mm; θ=0.00° |
| Shape 2 | 2 | Line | (-12.714, -17.500) | (-17.500, -2.772) | L=15.486 mm; θ=108.00° |
| Shape 2 | 3 | Line | (-17.500, -2.772) | (-17.500, 8.917) | L=11.688 mm; θ=90.00° |
| Shape 2 | 4 | Line | (-17.500, 8.917) | (-5.686, 17.500) | L=14.603 mm; θ=36.00° |
| Shape 2 | 5 | Line | (-5.686, 17.500) | (5.686, 17.500) | L=11.372 mm; θ=0.00° |
| Shape 2 | 6 | Line | (5.686, 17.500) | (17.500, 8.917) | L=14.603 mm; θ=144.00° |
| Shape 2 | 7 | Line | (17.500, 8.917) | (17.500, -2.772) | L=11.688 mm; θ=90.00° |
| Shape 2 | 8 | Line | (17.500, -2.772) | (13.538, -14.967) | L=12.823 mm; θ=72.00° |
| Shape 2 | 9 | Line | (13.538, -14.967) | (7.273, -17.500) | L=6.757 mm; θ=22.02° |
| Shape 2 | 10 | Line | (7.273, -17.500) | (-8.049, -17.500) | L=15.322 mm; θ=0.00° |
| Shape 3 | 1 | Line | (-9.873, 17.500) | (3.891, 17.500) | L=13.763 mm; θ=0.00° |
| Shape 3 | 2 | Arc | (3.891, 17.500) | (17.500, 3.891) | R=17.927 mm; sweep=64.93°; center= (0.000, 0.000) |
| Shape 3 | 3 | Line | (17.500, 3.891) | (17.500, 0.000) | L=3.891 mm; θ=90.00° |
| Shape 3 | 4 | Line | (17.500, 0.000) | (16.222, -6.765) | L=6.884 mm; θ=79.30° |
| Shape 3 | 5 | Line | (16.222, -6.765) | (17.500, -9.947) | L=3.430 mm; θ=111.88° |
| Shape 3 | 6 | Line | (17.500, -9.947) | (17.500, -13.298) | L=3.351 mm; θ=90.00° |
| Shape 3 | 7 | Line | (17.500, -13.298) | (2.362, -17.500) | L=15.710 mm; θ=15.51° |
| Shape 3 | 8 | Line | (2.362, -17.500) | (-9.801, -17.500) | L=12.163 mm; θ=0.00° |
| Shape 3 | 9 | Arc | (-9.801, -17.500) | (-17.500, -9.947) | R=8.357 mm; sweep=80.37°; center= (-18.121, -18.281) |
| Shape 3 | 10 | Line | (-17.500, -9.947) | (-17.500, -4.802) | L=5.145 mm; θ=90.00° |
| Shape 3 | 11 | Arc | (-17.500, -4.802) | (-17.500, 10.921) | R=10.309 mm; sweep=99.39°; center= (-24.169, 3.059) |
| Shape 3 | 12 | Line | (-17.500, 10.921) | (-17.500, 14.638) | L=3.717 mm; θ=90.00° |
| Shape 3 | 13 | Line | (-17.500, 14.638) | (-9.873, 17.500) | L=8.147 mm; θ=20.57° |
| Shape 4 | 1 | Line | (-8.049, 17.500) | (14.841, 10.206) | L=24.024 mm; θ=162.32° |
| Shape 4 | 2 | Line | (14.841, 10.206) | (17.500, 3.420) | L=7.288 mm; θ=111.40° |
| Shape 4 | 3 | Line | (17.500, 3.420) | (17.500, -7.761) | L=11.182 mm; θ=90.00° |
| Shape 4 | 4 | Line | (17.500, -7.761) | (2.730, -17.500) | L=17.692 mm; θ=33.40° |
| Shape 4 | 5 | Line | (2.730, -17.500) | (-14.141, -17.500) | L=16.871 mm; θ=0.00° |
| Shape 4 | 6 | Line | (-14.141, -17.500) | (-17.500, -4.408) | L=13.516 mm; θ=104.39° |
| Shape 4 | 7 | Line | (-17.500, -4.408) | (-17.500, -1.954) | L=2.454 mm; θ=90.00° |
| Shape 4 | 8 | Line | (-17.500, -1.954) | (-14.141, 17.500) | L=19.742 mm; θ=80.20° |
| Shape 4 | 9 | Line | (-14.141, 17.500) | (-12.714, 17.500) | L=1.427 mm; θ=0.00° |
| Shape 4 | 10 | Line | (-12.714, 17.500) | (-8.049, 17.500) | L=4.666 mm; θ=0.00° |
| Shape 5 | 1 | Line | (-2.245, -11.933) | (5.756, -16.315) | L=9.122 mm; θ=151.29° |
| Shape 5 | 2 | Arc | (5.756, -16.315) | (14.553, 14.097) | R=46.480 mm; sweep=39.82°; center= (-31.826, 11.034) |
| Shape 5 | 3 | Line | (14.553, 14.097) | (-0.406, 16.315) | L=15.122 mm; θ=171.57° |
| Shape 5 | 4 | Line | (-0.406, 16.315) | (-14.654, 12.360) | L=14.787 mm; θ=15.51° |
| Shape 5 | 5 | Arc | (-14.654, 12.360) | (-1.687, -9.690) | R=53.312 mm; sweep=27.76°; center= (-52.783, -24.900) |
| Shape 5 | 6 | Arc | (-1.687, -9.690) | (-2.245, -11.933) | R=7.427 mm; sweep=17.91°; center= (-9.086, -9.041) |

*__Table S2.__ Boundary coordinates and segment-level dimensions extracted from the supplied CAD files. A negligible duplicate drafting segment (~0.0005 mm) in Fabrication1.DXF is not part of the closed Shape 5 boundary and is omitted.*

**Electrode, attachment, and interconnection information**

**Electrode coverage:** The devices are fabricated from 0.5-mm-thick $LiNbO_3$ with nominal 1.0-µm aluminum deposited on the top and bottom faces before laser definition of the lateral outline. Thus, the outline CAD defines the $LiNbO_3$/electrode plate boundary, while the electrode coverage is nominally the full top and bottom plate area rather than a patterned lateral electrode geometry.

**Attachment and wire bonds:** The DXF files contain only the resonator outlines and do not specify PCB attachment regions or wire-bond coordinates; these packaging parameters are therefore reported separately from the resonator geometry. The bond counts and attachment conditions for the shape-comparison devices are summarized in Table S3. All devices compared in Fig. 3 use identical interconnections, with 3 wire bonds on both the top and bottom electrodes, thereby holding the wire-bond configuration constant across the Fig. 3 comparison. However, active area and area-to-perimeter ratio still differ among the tested outlines; Fig. 3 is therefore interpreted as a comparison among fabricated geometries rather than as a strict one-variable isolation of lateral shape. Wire bonding is examined separately in Fig. 4, where Device A uses 3 wire bonds on each electrode side and Device B uses 20 wire bonds on each side.

| **Device** | **$LiNbO_3$ thickness (mm)** | **Electrode coverage** | **Attachment** | **No. (Top)** | **No. (Back)** |
|---|---|---|---|---|---|
| Square | 0.50 | Blanket Al, top and bottom | PCB-mounted adhesive | Fig.3:3 | Fig.3:3 |
| Shape 1 | 0.50 | Blanket Al, top and bottom | PCB-mounted adhesive | Fig.3:3 | Fig.3:3 |
| Shape 2 | 0.50 | Blanket Al, top and bottom | PCB-mounted adhesive | Fig.3:3 | Fig.3:3 |
| Shape 3 | 0.50 | Blanket Al, top and bottom | PCB-mounted adhesive | Fig.3:3 | Fig.3:3 |
| Shape 4 | 0.50 | Blanket Al, top and bottom | PCB-mounted adhesive | Fig.3:3 | Fig.3:3 |
| Shape 5 | 0.50 | Blanket Al, top and bottom | PCB-mounted adhesive | Fig.3/Fig4A:3<br>Fig.4B:20 | Fig.3/Fig4A:3<br>Fig.4B:20 |

*__Table S3.__ Geometry-associated fabrication and packaging information.*

**Uniform scaling of shape 5 for the area-scaling study**

**Scaling rule**

All devices in the Fig. 5 area-scaling study use the same normalized Shape 5 boundary. Starting from the 35/35 reference CAD, the entire lateral boundary was uniformly (isotropically) scaled about its center using linear scale factors s = 10/35, 15/35, 20/35, 25/35, and 35/35. No independent change was made to taper angles, arc sweep angles, concavity, or normalized boundary shape. This scaling strategy isolates the effect of overall lateral size/area while preserving the same degree of geometric asymmetry used to disrupt coherent lateral standing-wave conditions.

**Geometric consequences of uniform scaling and wire bonding**

For a uniform scale factor s, all Cartesian boundary coordinates, straight-edge lengths, overall width/height, perimeter, and curvature radii are multiplied by s, while area is multiplied by $s^2$. Straight-edge orientations and arc sweep angles remain unchanged. Thus, the complete CAD for each area-scaled device can be reproduced directly from the Shape 5 boundary definition in Table S4 using (x, y) → s(x, y) and R → sR.

| Linear scale | Nominal Area ($mm^2$) | Nominal Perimeter (mm) | No. (Top) | No. (Back) |
|---|---|---|---|---|
| 10/35 | 40 | 30 | 20 | 20 |
| 15/35 | 100 | 40 | 20 | 20 |
| 20/35 | 190 | 60 | 20 | 20 |
| 25/35 | 290 | 70 | 20 | 20 |
| 35/35 | 560 | 100 | 20 | 20 |

***Table S4.*** *Nominal area/perimeter values and interconnection conditions of the uniformly scaled Shape 5 devices used in Fig. 5.*

**Transferability across size**

Because the same normalized Shape 5 contour is retained, this series provides a direct size-scaling test of the boundary-design concept. The principal TS resonance remains primarily set by plate thickness, whereas changing the lateral scale changes plate area, capacitance/impedance level, perimeter-to-area ratio, and measured $Q$. The results therefore support using the asymmetric boundary as a scalable design template whose overall size can be selected according to the electrical impedance and power-converter requirements, while the normalized asymmetry is preserved.

**Area-dependent effective inductance and impedance range**

To connect the area-scaling study more directly to converter-relevant reactive impedance, the measured two-port S-parameters of the five uniformly scaled Shape-5 devices were reprocessed using the same 50 Ω S to Y conversion used elsewhere in this work. The transfer-branch series impedance was obtained from $Z_{series} = -1/Y_{21}$. Within the uninterrupted interval between the $Y_{21}$-defined series resonance $f_s$ and parallel resonance $f_p$ for which $\mathrm{Im}(Z_{series}) > 0$, the resonator presents an inductive series reactance. A frequency-dependent effective series inductance was therefore defined as $L_{eff}(f) = \mathrm{Im}[Z_{series}(f)]/(2\pi f)$. Because $\mathrm{Im}(Z_{series})$ approaches zero at series resonance, the lower bound of the reported effective-inductance range is approximately zero; the upper bound is the maximum $L_{eff}$ reached before the inductive interval terminates. All values are extracted from the as-measured packaged-device data without PCB de-embedding, consistent with the application-relevant measurements reported in the main manuscript.

The area series reveals a size–impedance–loss trade-off rather than a monotonic optimum. Across the 40–560 mm² range, reducing area increases the accessible maximum effective inductance but also increases series resistance and lowers $Q_s$. Specifically, the 40 mm² device reaches approximately 68.7 µH but has $R(f_s)$ = 37.54 Ω and $Q_s$ = 90, whereas the 560 mm² device reaches approximately 35.9 µH with $R(f_s)$ = 0.225 Ω and $Q_s$ = 1146. The uninterrupted inductive bandwidth also increases from 513 to 723 kHz as area increases. The 560 mm² device therefore represents the low-resistance/high-*Q* endpoint of the measured design space rather than a universally preferred footprint. For a specific DC–DC converter, the lateral scale can be selected according to the required reactive-impedance/effective-inductance range, loss, load range, and operating frequency while retaining the same normalized asymmetric Shape-5 boundary.

| Area (mm²) | $f_s$ (MHz) | $f_p$ (MHz) | 3-dB $Q_s$ | $R(f_s)$ (Ω) | Continuous inductive region (MHz) | Inductive BW (kHz) | Effective $L_{eff}$ range (µH) |
|---|---|---|---|---|---|---|---|
| 40 | 3.681 | 4.210 | 90 | 37.54 | 3.681–4.193 | 513 | ≈0–68.7 |
| 100 | 3.689 | 4.329 | 279 | 7.02 | 3.689–4.292 | 603 | ≈0–53.9 |
| 190 | 3.673 | 4.346 | 353 | 3.00 | 3.673–4.319 | 646 | ≈0–40.6 |
| 290 | 3.665 | 4.352 | 495 | 1.01 | 3.665–4.352 | 687 | ≈0–49.9 |
| 560 | 3.667 | 4.391 | 1146 | 0.225 | 3.667–4.390 | 723 | ≈0–35.9 |

***Table S5.*** *Measured area-dependent electrical design range of the uniformly scaled Shape-5 resonators.*

The non-monotonic maximum Leff across area (for example, the 190 and 290 mm² devices) shows that the measured response does not follow a simple inverse-area scaling law. The reported values reflect the complete two-port resonator response together with package/interconnection loading and are therefore treated as experimentally measured design ranges rather than a universal area-scaling relation.

Area also changes the capacitive and motional impedance scales. Because the devices use the same 0.5-mm $LiNbO_3$ thickness and nominal full-face electrodes, increasing lateral area is expected to increase the static capacitance and reduce the electrical impedance level, although the measured response can also include electrode, contact, and package parasitics. For the representative 560 mm² device, the MBVD fit reported in the main manuscript gives $C_0$ = 486 pF, Lm = 8.9 μH, Cm = 211 pF, and Rm = 0.155 Ω. Thus, resonator area should be selected jointly with capacitance, motional impedance, effective-inductance range, load range, and operating frequency rather than optimized independently.

Although the 560 $mm^2$ resonator has a relatively large lateral footprint, the active $LiNbO_3$ plate is only 0.5 mm thick, corresponding to an active material volume of approximately 280 $mm^3$ (0.28 $cm^3$). For volumetric context, a representative commercial 33-μH high-current shielded power inductor (Coilcraft SER2918H-333KL) has maximum package dimensions of approximately 27.94 × 27.9 × 17.78 $mm^3$, corresponding to a geometric package envelope of approximately 13.9 $cm^3$. On this limited active-plate-versus-package-envelope basis, the piezoelectric plate is approximately 50 times smaller in volume. This comparison is intended only to illustrate the volumetric potential of the thin piezoelectric element and is not a claim of direct component or converter-level equivalence. The magnetic inductor and piezoelectric resonator differ in operating principle, current handling, energy storage, large-signal power capability, thermal limits, and frequency dependence. In addition, the present PCB and wire-bond assembly was designed for laboratory characterization rather than minimum footprint. We therefore do not claim a 50× reduction in total packaged component volume; minimizing the complete footprint requires further resonator–package–converter co-design.

Expected converter power density is likewise not inferred from the present small-signal S-parameter measurements. The source-level sweep provides an excitation-sensitivity check over the measured range, but a defensible converter-level power-density estimate requires sustained large-signal characterization of resonator voltage and current, transferred and dissipated power, self-heating, nonlinear loss, contact reliability, thermal limits, and the area/volume of the complete converter implementation. The present results therefore establish the resonator-level area–impedance–loss trade-off, while converter-level power density remains a subject for future system-level validation.

## Fabrication and Measurement Setup

This supplementary information provides additional fabrication, measurement, and device photographs supporting the main manuscript.

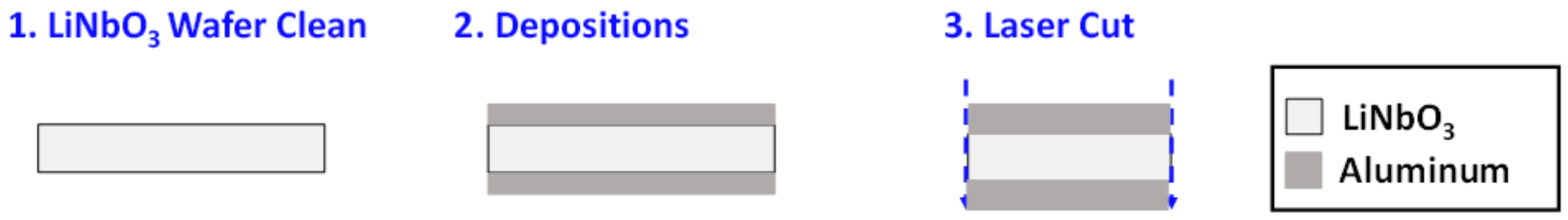

***Fig. S2.*** *Lithography-free fabrication process flow. The process consists of LiNbO₃ wafer preparation, blanket electrode deposition, and laser cutting to define the resonator body without photolithographic patterning.*

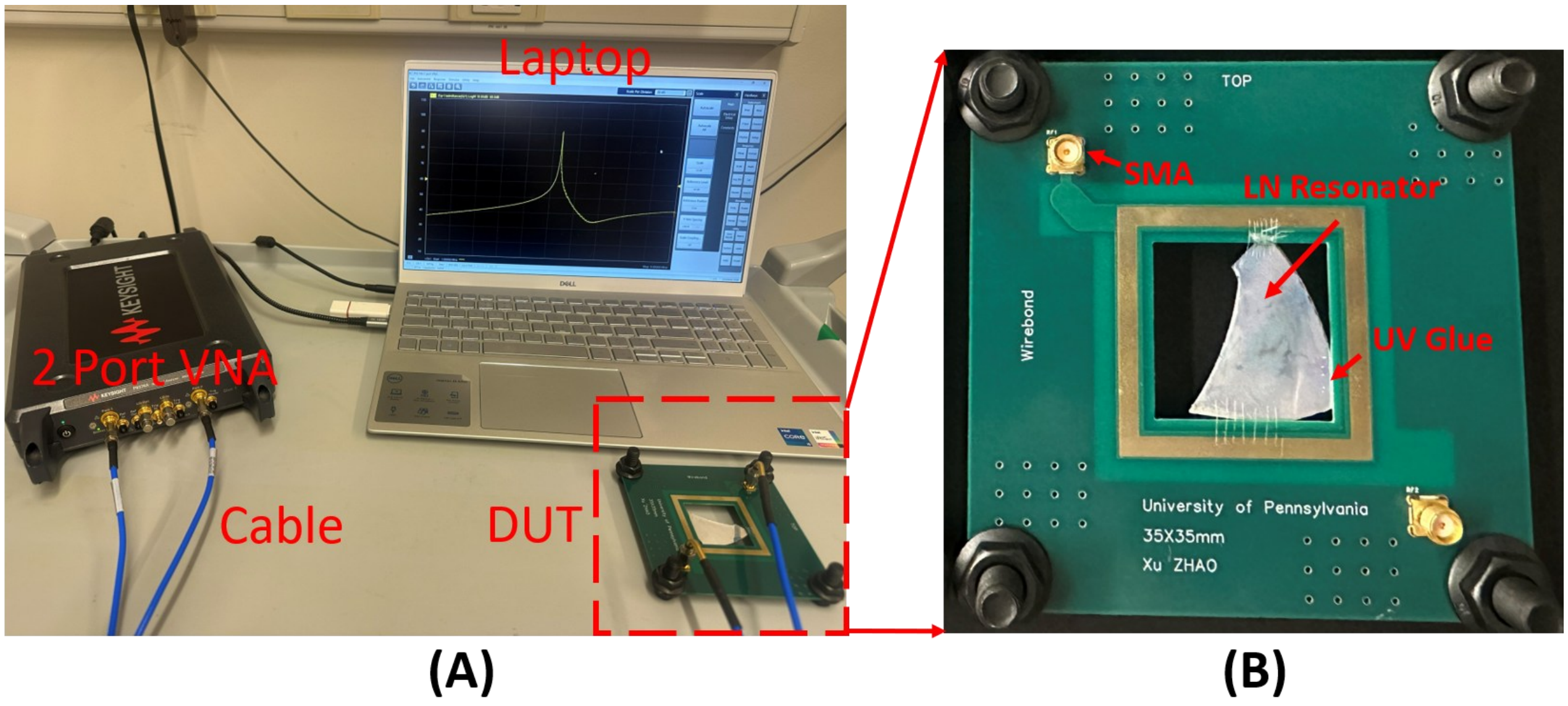

***Fig. S3.*** *(A) Measurement setup for the packaged thickness-shear LiNbO₃ resonator. (B) The device was mounted on a printed circuit board and characterized using a two-port vector network analyzer after calibration.*

## Wirebond Details

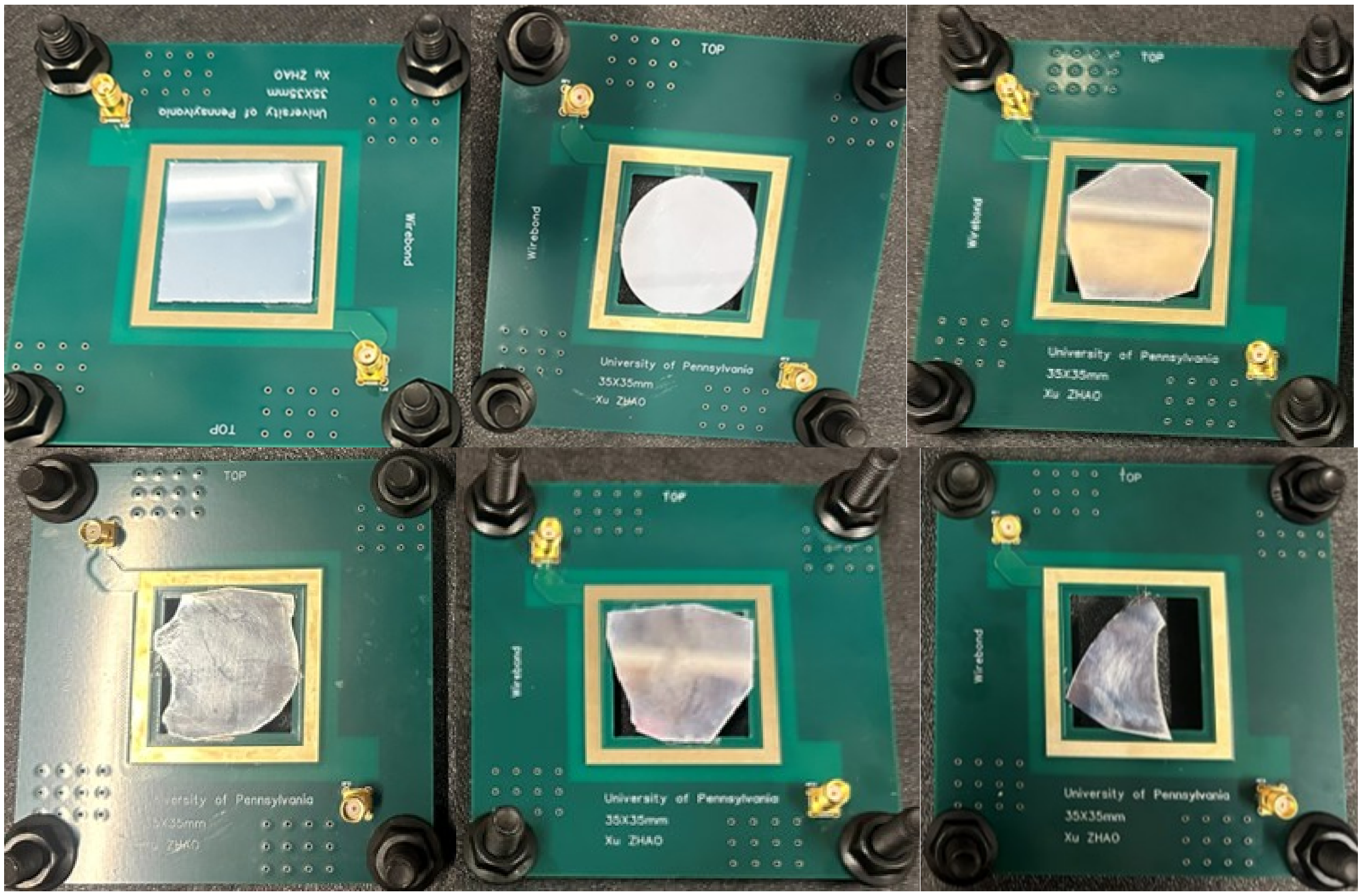

*Fig. S4. Photographs of the PCB-mounted square reference and selected shaped resonators in Fig. 3.*

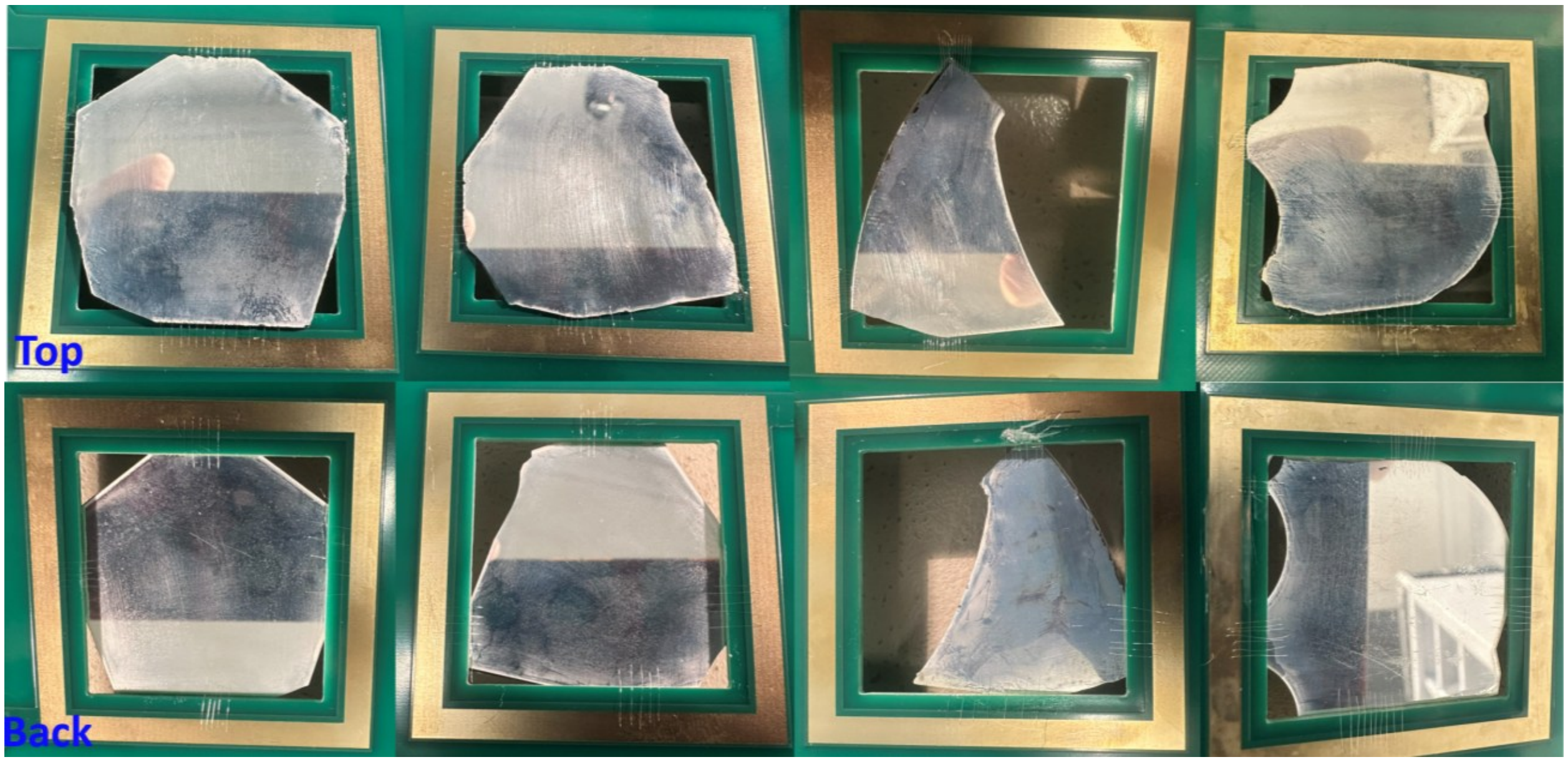

*Fig. S5. Additional top- and bottom-side photographs of packaged resonators used to evaluate interconnection configurations in Fig. 4.*

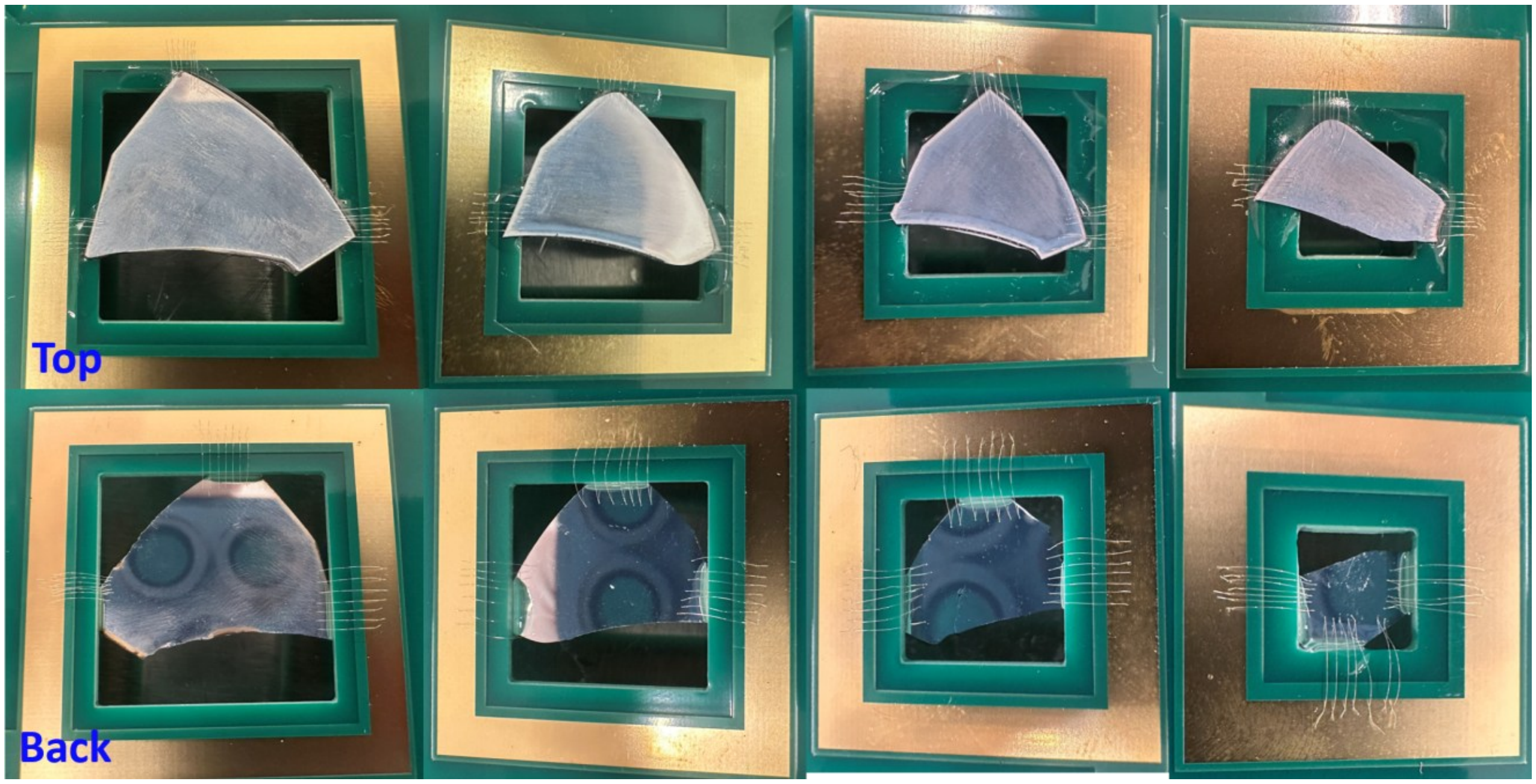

***Fig. S6.*** *Photographs of the top and back sides of resonators with different shaped outlines used for spurious-mode suppression studies.*

## Calibration and Measurement Details

This section describes the calibration plane and verification, measurement repeatability, fixture/de-embedding sensitivity, fabrication repeatability, and excitation-level dependence of the packaged $LiNbO_3$ thickness-shear resonator.

**Calibration and measurement configuration**

Measurements were performed with a Keysight P9374A vector network analyzer in a 50 Ω environment. A two-port short-open-load-through calibration was performed to the ends of the measurement cables before each measurement session. The calibration plane is therefore located at the cable ends. Unless explicitly noted as de-embedded below, the reported spectra include the PCB connectors, routing, contact pads, wire bonds, and package. The fitted resistance values are interpreted as effective parameters of the measured packaged response rather than as fixture-independent intrinsic material values.

For quality-factor extraction, the measured S-parameter matrix was converted to Y parameters using MATLAB where s2y (S Parameters, 50), and $Y_{21}$ = Y (2,1) was converted to dB as $20\log_{10}(|Y_{21}|)$. The series and parallel resonances were taken at the maximum and minimum of $Y_{21}$, respectively. $Q_s$ was calculated as $f_s$ divided by the 3-dB bandwidth defined by the lower- and upper-frequency 3-dB boundaries were taken as the nearest sampled frequencies on either side of $f_s$ at which $|Y_{21}|$ across the $Y_{21, max} - 3$ dB level, exactly as implemented in the supplied MATLAB code. For impedance and resistance, the same full two-port S-to-Y conversion was used, and the series-path impedance was calculated as $Z_{series} = -1/Y_{21}$, with $R = Re(Z_{series})$. This formulation is appropriate for the transfer branch because $Y_{21}$ is the negative series-branch admittance and, unlike an $S_{21}$-only conversion, it retains the influence of the complete measured two-port S matrix.

**Open short load and thru result**

The measured thru response is close to unity transmission over the band: the mean S21 magnitude is 0.002 dB (range -0.002 to 0.005 dB). The maximum S11 magnitude is -54.5 dB. This check verifies the cable-end calibration, while not replacing device-level fixture de-embedding.

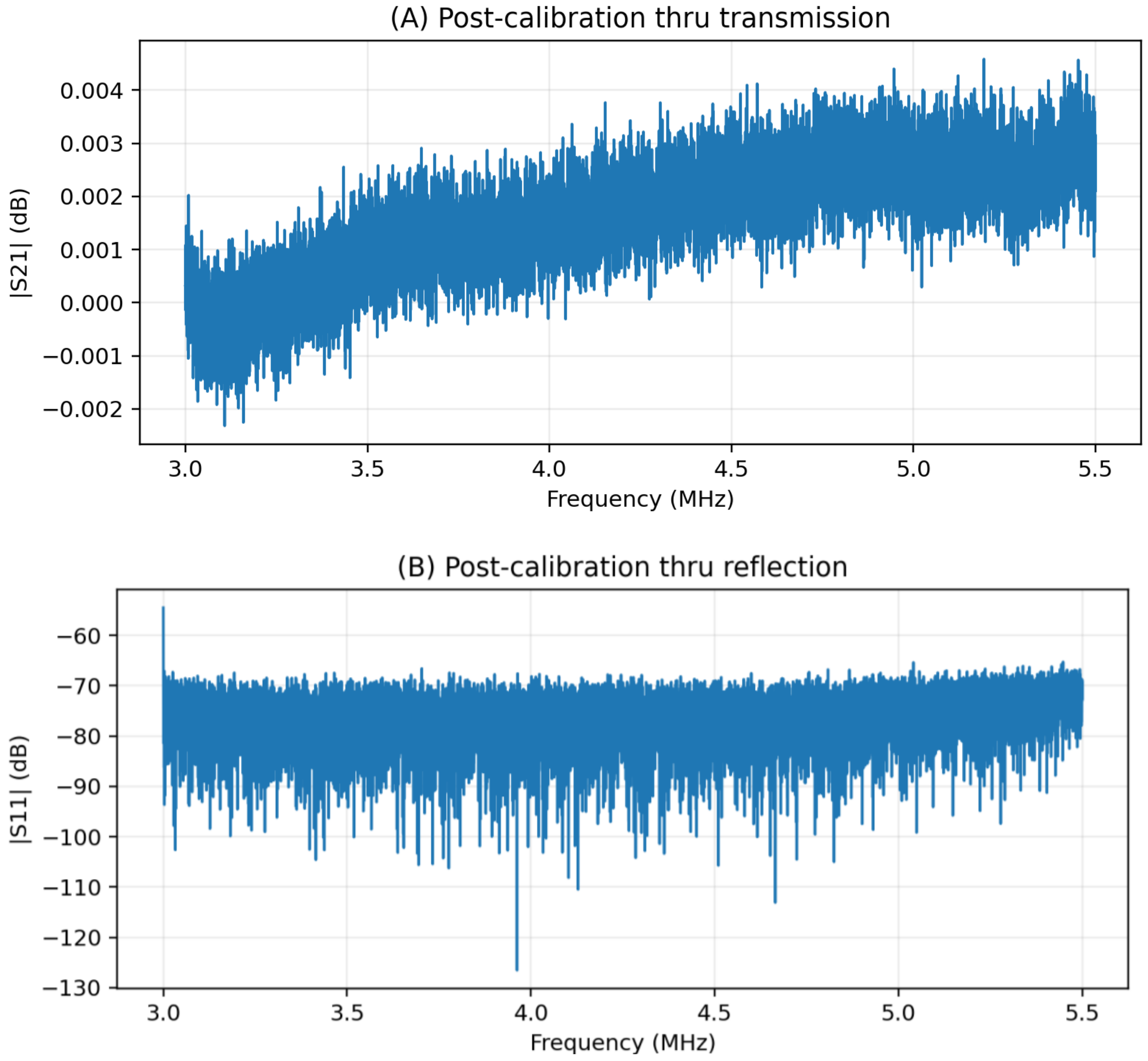

***Fig. S7.*** *Post-calibration thru verification over the 3.0–5.5 MHz: (A) $S_{21}$ transmission and (B) $S_{11}$ reflection*

**Repeatability with independent calibrations**

The same packaged device was remeasured in three independent sessions on June 5, August 4, and August 9, 2026. A new cable-end calibration was performed for each session. The purpose of this test is to determine whether the resonance position and overall spectral features persist when calibration is repeated at different times. The raw $S_{21}$ and $S_{11}$ responses are plotted separately so that the transmission resonance and the reflection behavior can be assessed.

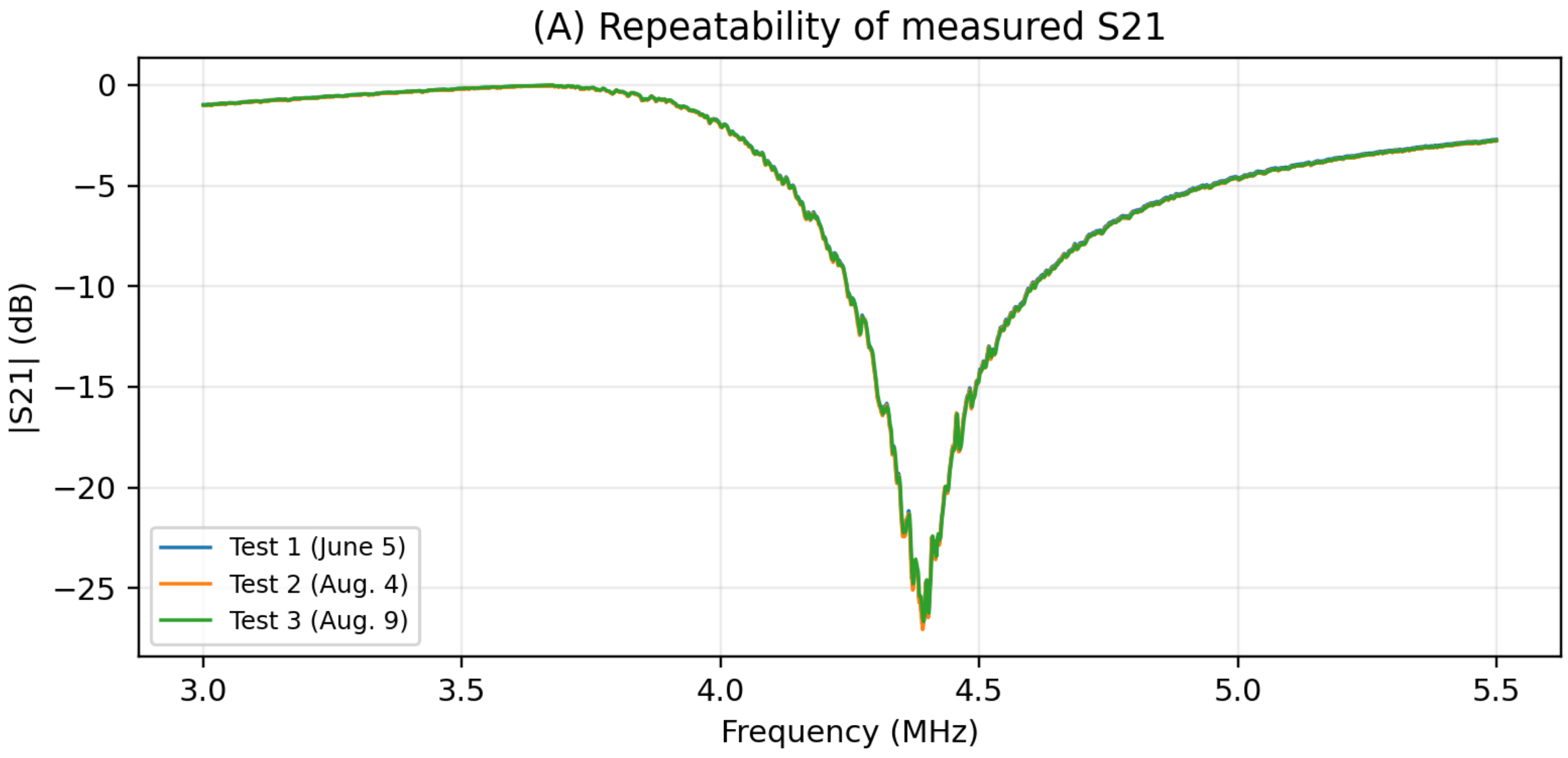

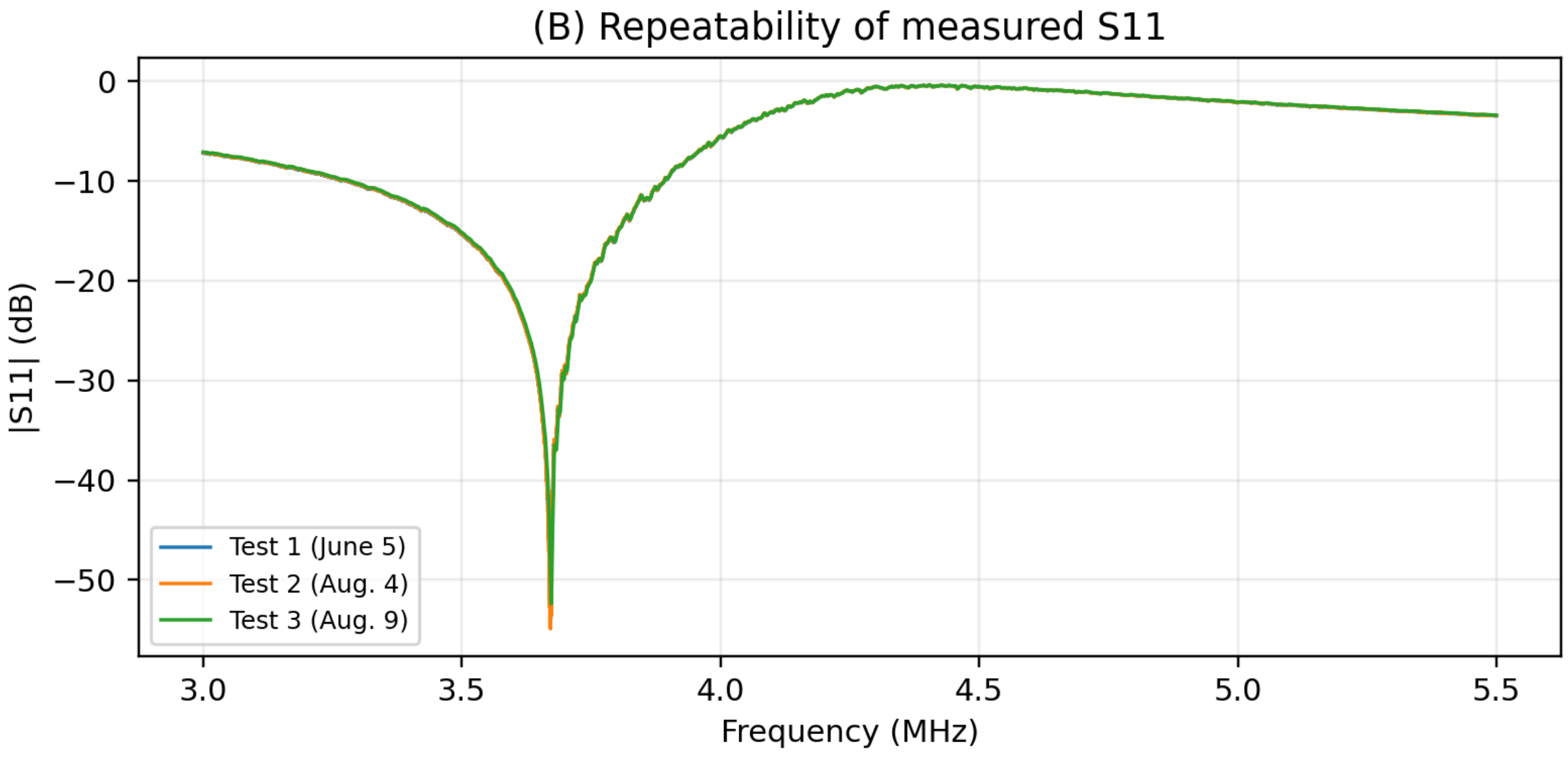

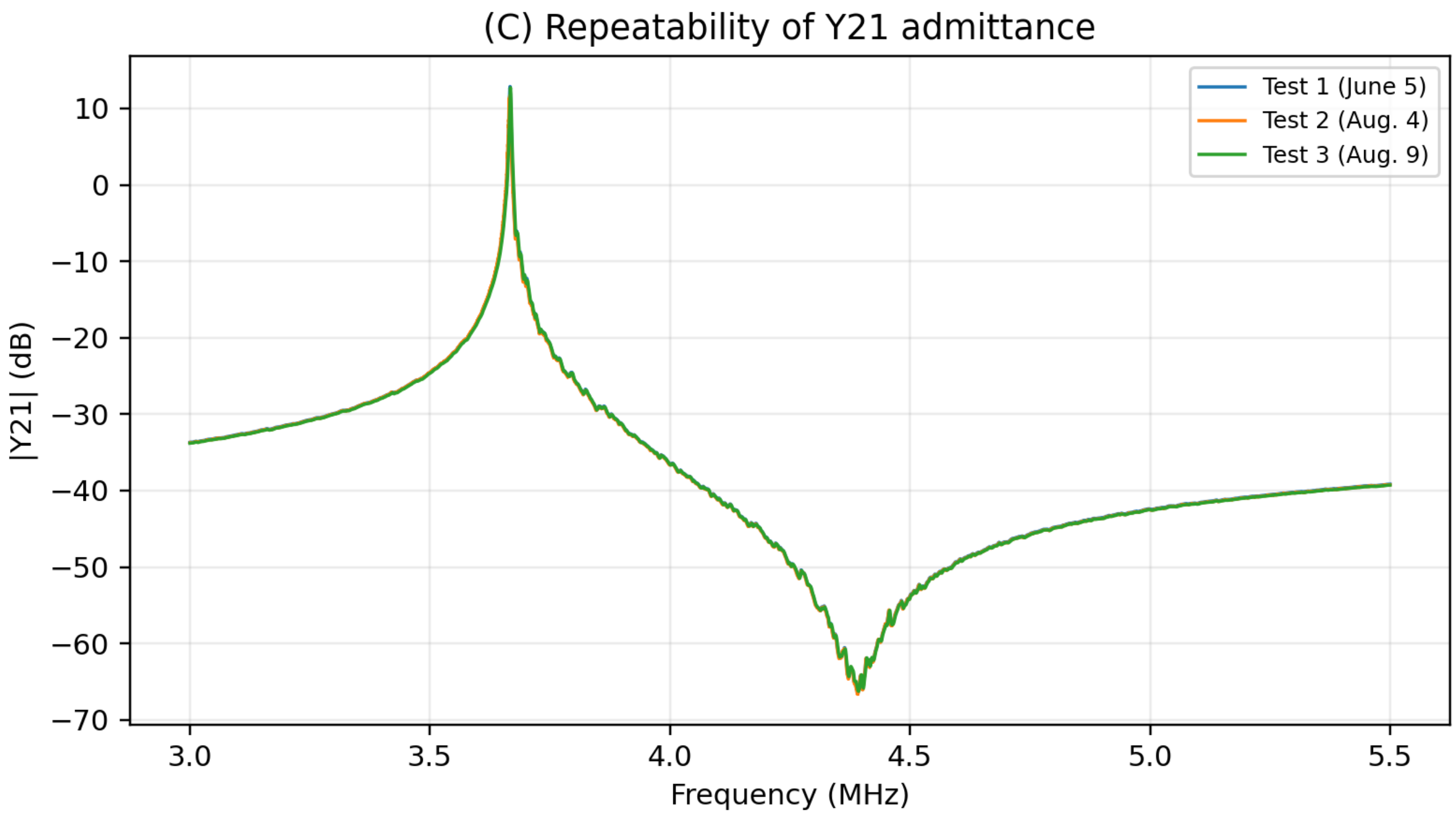

***Fig. S8.*** *Repeatability of the same packaged resonator measured in three independent sessions: (A) $S_{21}$, (B) $S_{11}$, and (C) $Y_{21}$ admittance.*

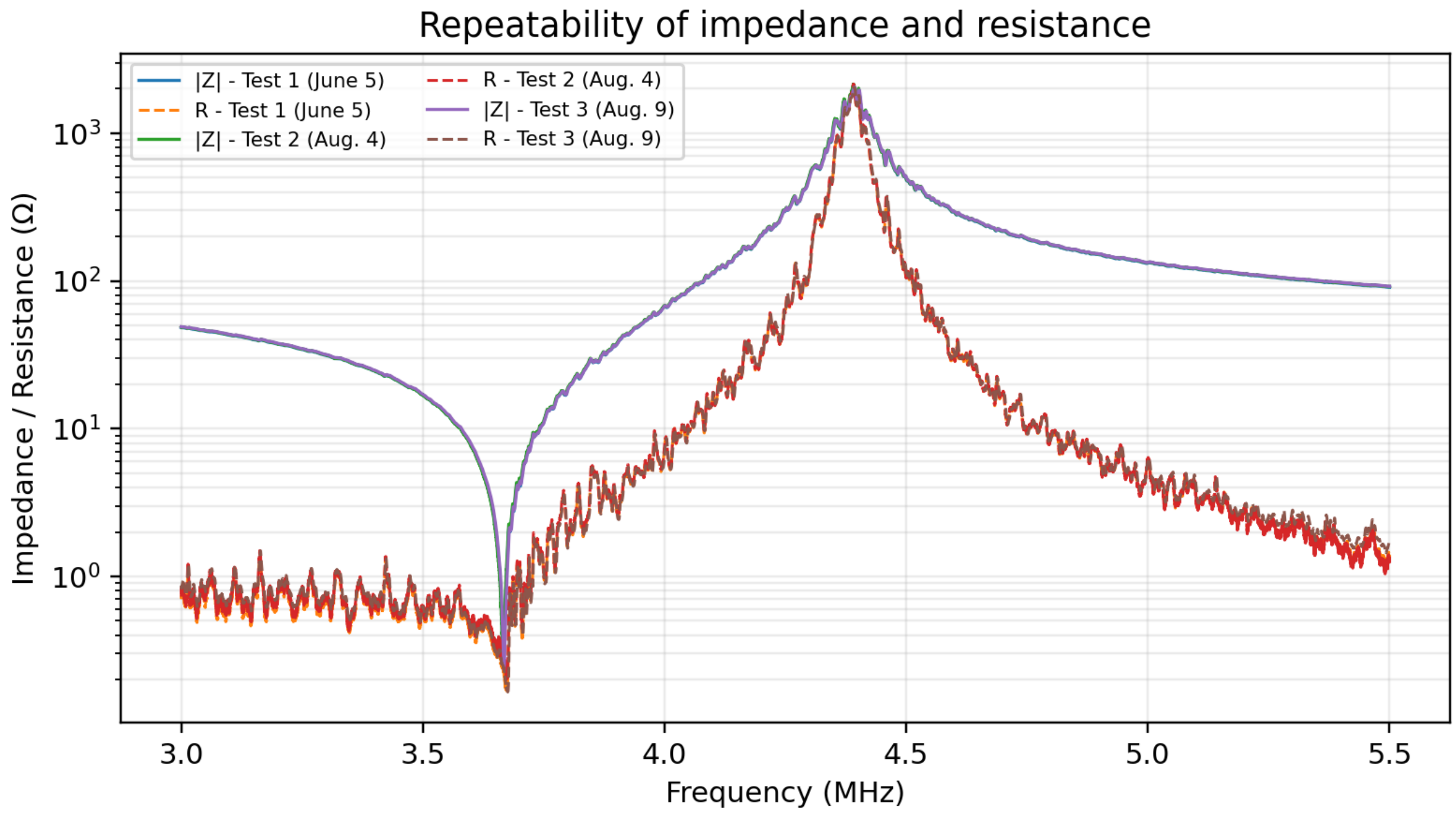

***Fig. S9.*** *Impedance magnitude and resistance of the same packaged resonator measured in three independent sessions.*

| Measurement | $f_s$ (MHz) | $f_p$ (MHz) | $k_t^2$ (%) | 3-dB $Q_s$ |
|---|---|---|---|---|
| Test 1 (June 5) | 3.666953 | 4.391468 | 53.567 | 1146 |
| Test 2 (Aug. 4) | 3.666327 | 4.390756 | 53.570 | 1183 |
| Test 3 (Aug. 9) | 3.667927 | 4.392656 | 53.569 | 1111 |

***Table S6.*** *Measured resonator parameters for the same physical resonator device measured in 3 different measurement sessions. The characteristics of the resonator and associated PCB and wirebond parasitics are highly repeatable across measurement sessions.*

$Q_s$, $f_s$, $f_p$, $k_t^2$, and the corresponding FoM values were reprocessed directly from $Y_{21}$ using the exact MATLAB workflow described above (s2y conversion followed by the $Y_{21}$ 3-dB extraction). R at $f_s$ is obtained from $R = Re(-1/Y_{21})$ at the corresponding $Y_{21}$-derived $f_s$.

For the three repeated measurements, $|S_{11}|$ at the $Y_{21}$-defined series resonance is approximately -43.7, -42.7, and -42.4 dB, respectively. Thus, $S_{11}$ is not close to 0 dB at fs, where the 3-dB $Q_s$ is extracted. As expected for this two-port resonator, $S_{11}$ approaches 0 dB near the antiresonance (approximately -0.38 dB at its local maximum), which is shown explicitly in the separately plotted $S_{11}$ traces.

The resonance frequencies are highly repeatable: $f_s$ = 3.667069 MHz with a standard deviation of 0.806 kHz, and $f_p$ = 4.391626 MHz with a standard deviation of 0.960 kHz. The corresponding coupling coefficient varies by less than 0.003 percentage points. Using the same MATLAB $Y_{21}$ extraction, $Q_s$ = 1147 ± 36 (mean ± one standard deviation), with individual values ranging from 1111 to 1183. This repeatability supports the robustness of the 3-dB $Q_s$ extraction, while the independently extracted series-path resistance is also repeatable across the three sessions. The local maximum Bode $Q$ is treated separately because it is more sensitive to numerical phase differentiation and point-to-point measurement noise.

**PCB fixture de-embedding using a symmetric thru dummy**

To quantify the contribution of the PCB fixture, we measured a dedicated PCB thru dummy in which the resonator site was replaced by a direct electrical connection. The dummy used the same SMA connectors, PCB stack-up, transmission-line routing, and contact-pad geometry as the resonator fixture. The measured dummy response is close to an ideal thru over 3.0-5.5 MHz (mean $S_{21}$ = -0.0106 dB; $S_{21}$ = -0.0086 dB and $S_{11}$ = -48.0 dB near 3.67 MHz), allowing the repeatable PCB fixture contribution to be separated from the DUT response.

The PCB de-embedding follows the symmetric two-port fixture-removal method described by Wartenberg [1]. For the measured PCB thru dummy, the two identical half-fixture networks are obtained from the measured thru S-parameters as:

$$S_{11,p} = S_{22,p} = (S_{11} + S_{22})/(2 + S_{21} + S_{12}) \tag{S1}$$

$$S_{12,p} = S_{21,p} = (\sqrt{0.5(S_{12} + S_{21})(1 - S_{11,p}{}^{2})} \tag{S2}$$

These terms define the symmetric half-fixture scattering matrix $S_p$. A phase-continuous square-root branch is selected across frequency. The extracted $S_p$ and the measured DUT S-matrix are then converted to wave-cascade (T) matrices. The fixture-corrected DUT matrix is calculated as:

$$T_{\mathrm{DUT}} = T_p^{-1} T_{\mathrm{meas}} T_p^{-1} \tag{S3}$$

The resulting $T_{\mathrm{DUT}}$ is converted back to S-parameters and then to Y-parameters using the same 50-ohm reference impedance used in the manuscript analysis. Admittance, series impedance, resistance, and the 3-dB $Q_s$ are subsequently re-extracted using the same processing workflow as for the non-de-embedded data.

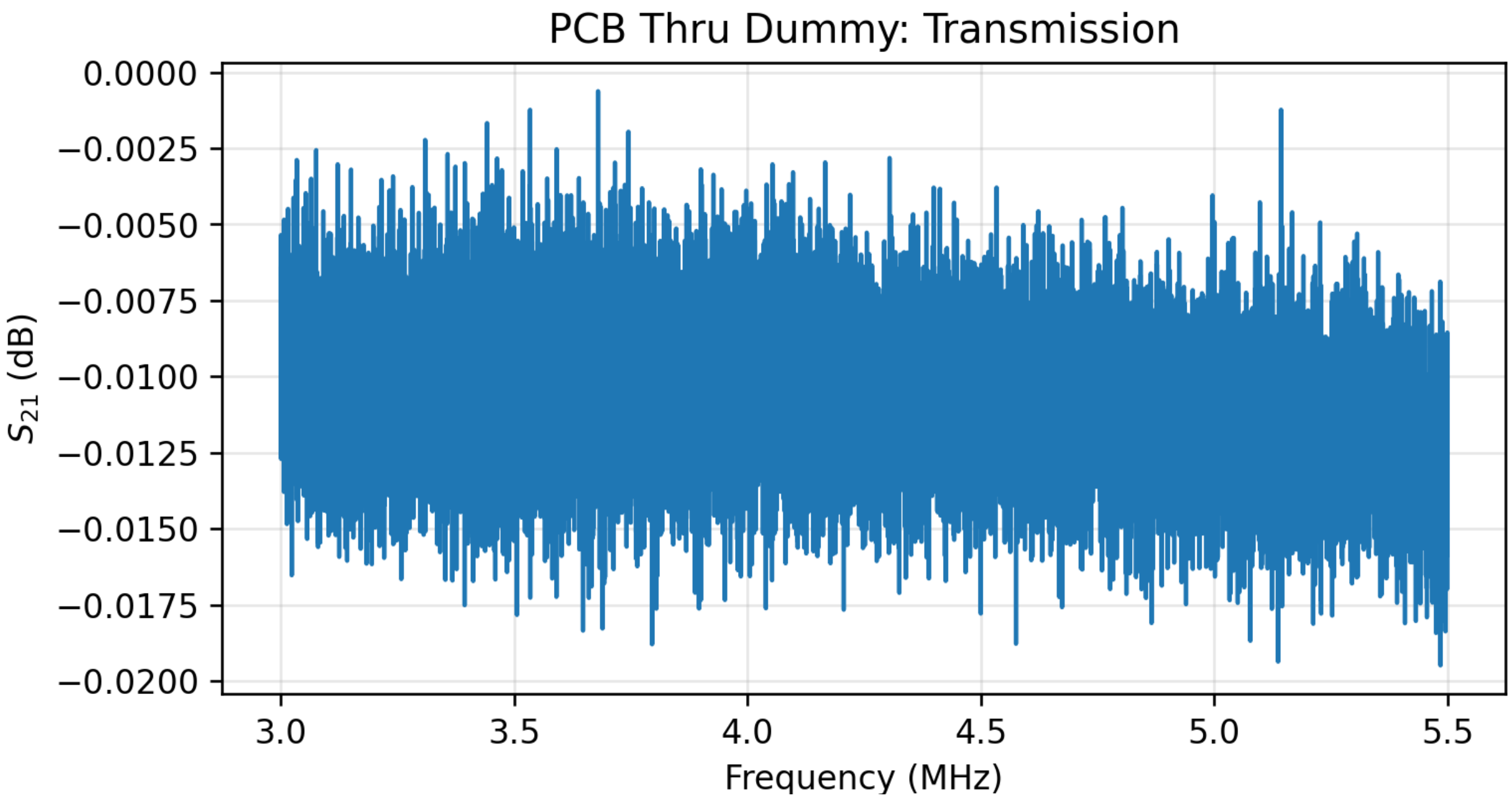

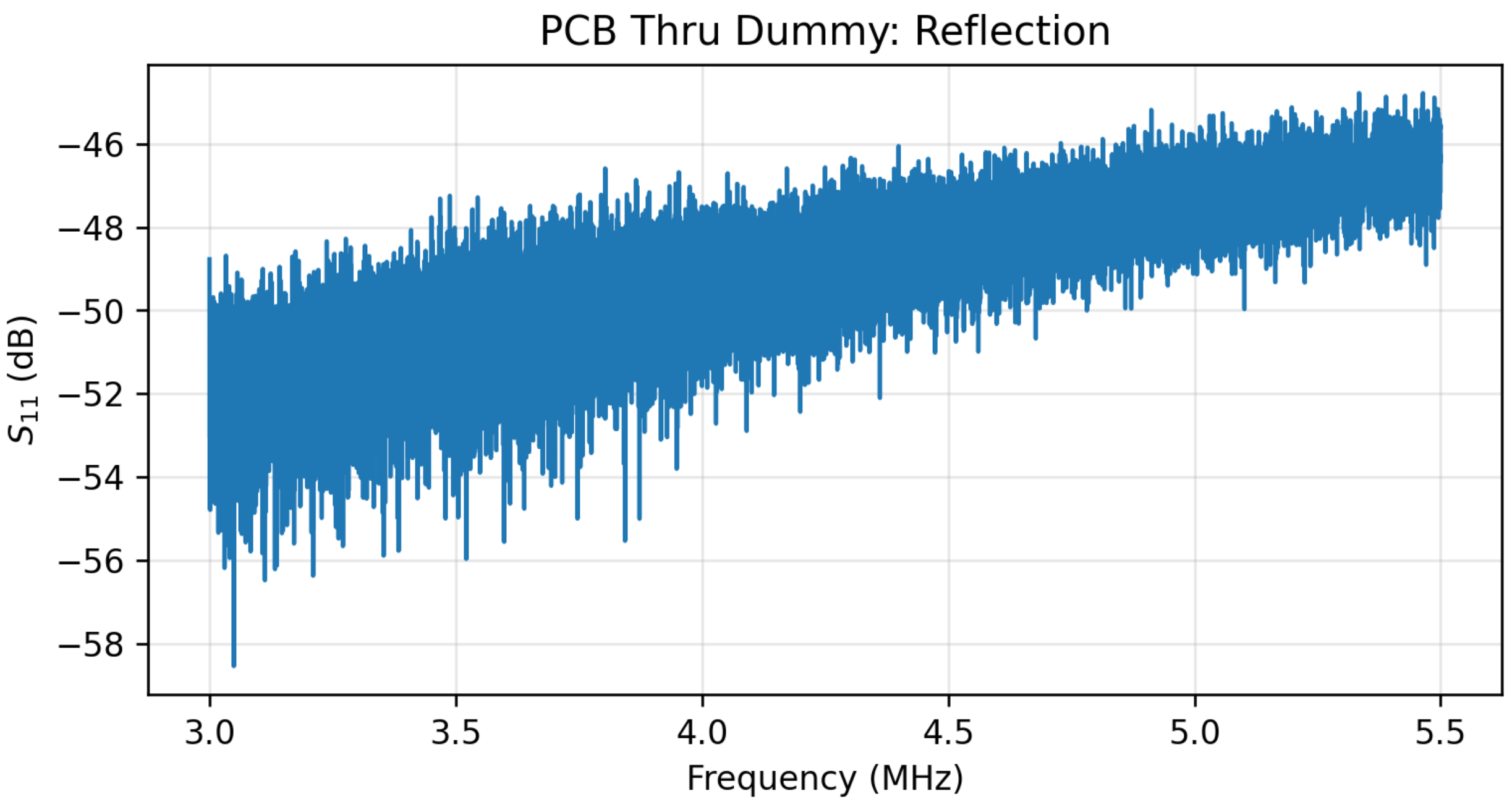

***Fig. S10.*** *Measured PCB thru dummy used for PCB-level de-embedding. $S_{21}$ transmission and $S_{11}$ reflection are plotted separately for clarity.*

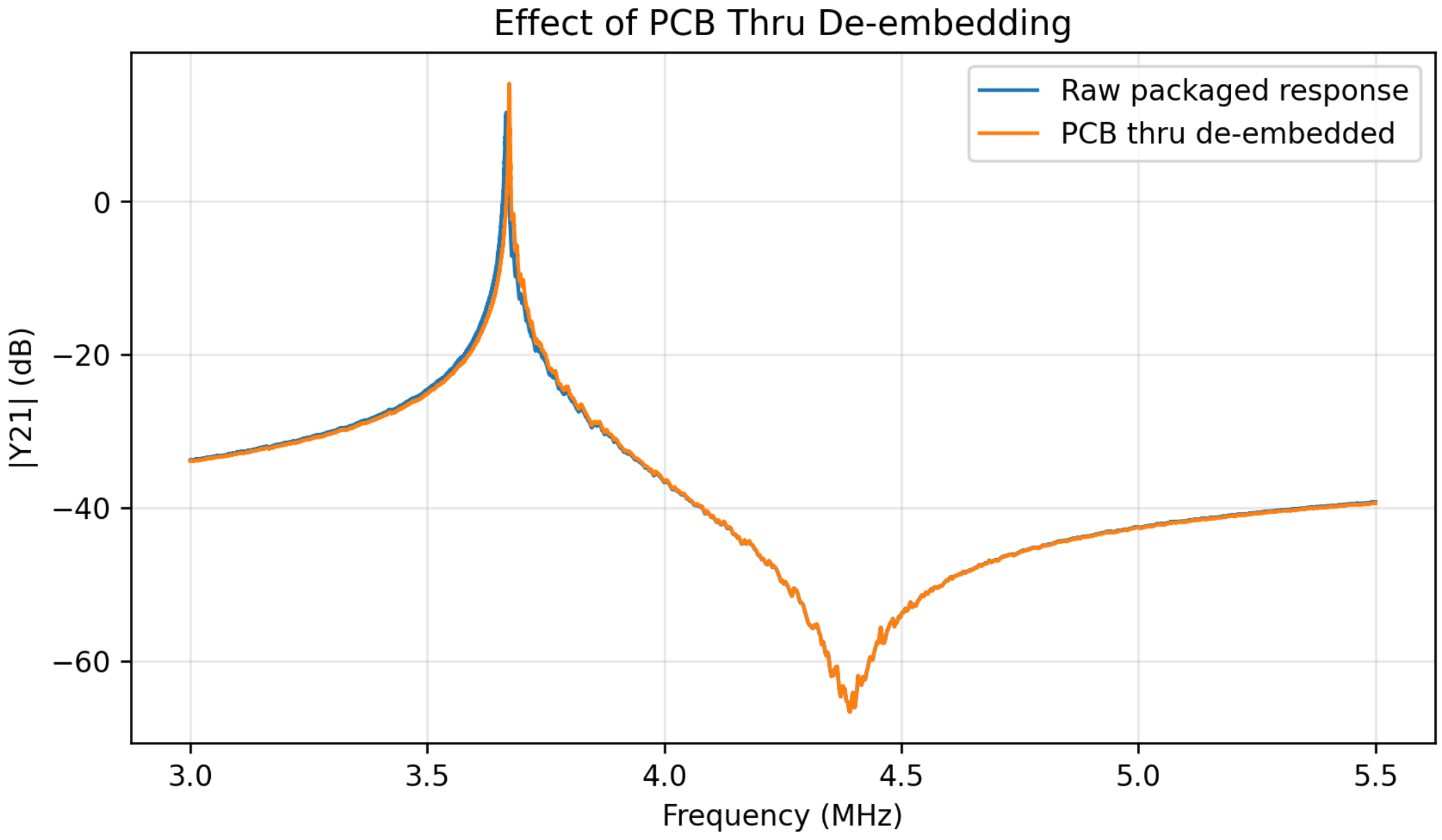

***Fig. S11.*** *Admittance before and after PCB-level thru de-embedding.*

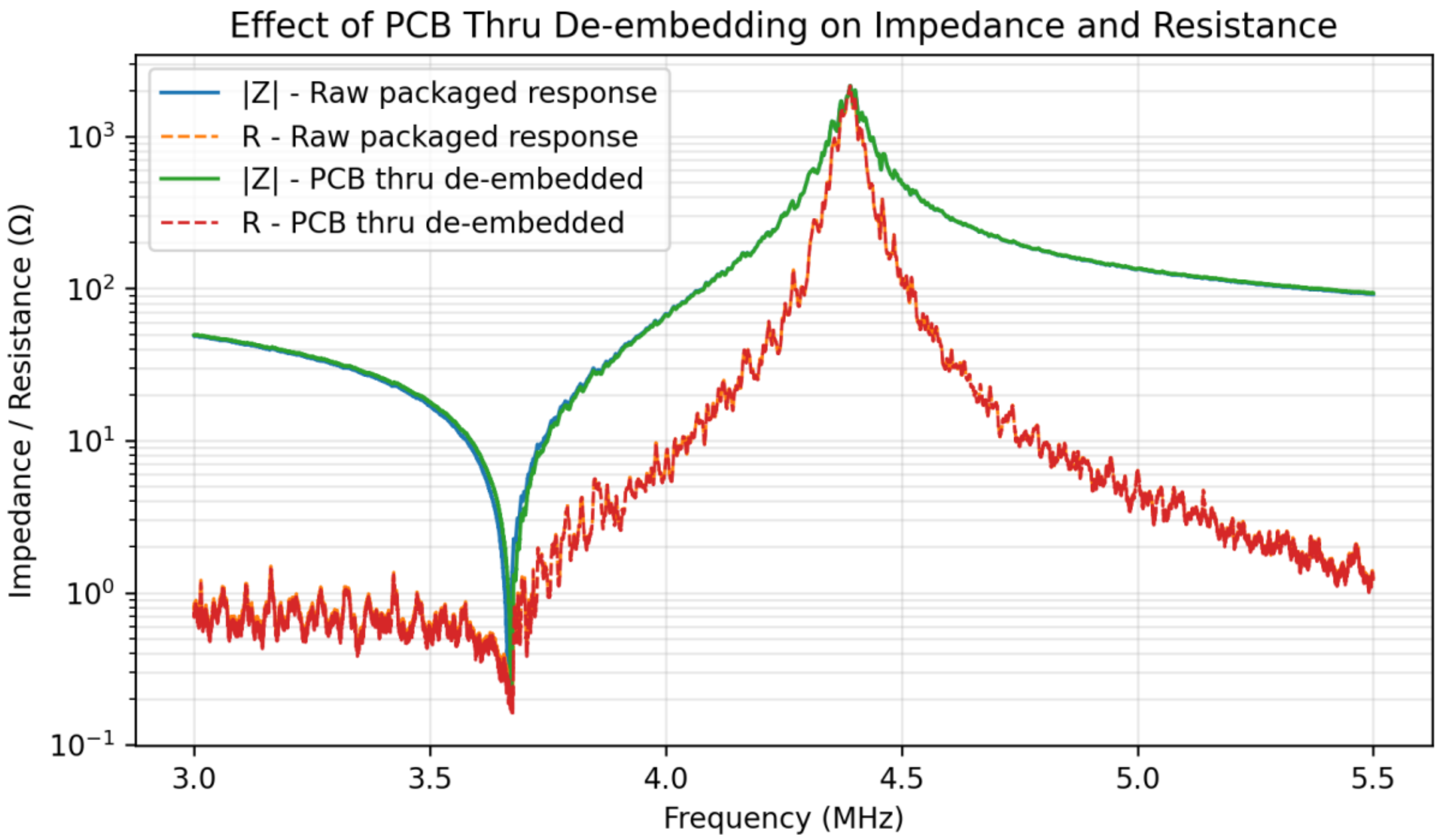

***Fig. S12.*** *Impedance magnitude and resistance before and after PCB-level thru de-embedding.*

| **Condition** | $f_s$ **(MHz)** | $f_p$ **(MHz)** | **3-dB** $Q_s$ | **R(**$f_s$**) (Ω)** |
|---|---|---|---|---|
| Raw packaged response | 3.67 | 4.39 | 1183 | 0.257 |
| PCB thru de-embedded | 3.67 | 4.39 | 2295 | 0.166 |

***Table S7.Measured*** *resonator parameters for the same physical device with and without PCB de-embedded.*

The PCB-level de-embedding changes the extracted loss-related quantities while preserving the overall resonance topology. Using the same $Y_{21}$-based MATLAB extraction, $Q_s$ changes from 1183 for the as-measured packaged response to 2295 after PCB de-embedding, while $R$ at fs changes from approximately 0.257 ohm to 0.166 ohm. This comparison demonstrates that the PCB fixture contributes appreciably to the measured loss and should be considered when interpreting intrinsic device parameters. However, the main manuscript intentionally retains the original non-de-embedded packaged-device results. The resonator is intended to operate wire-bonded on a PCB in a practical DC-DC converter, so the PCB routing, pads, contacts, and associated interconnection loss are part of the application-relevant electrical response seen by the converter. Accordingly, the headline values in the main text are explicitly identified as measured packaged device performance, while the PCB de-embedded results are provided here as a fixture-sensitivity and intrinsic device cross-check. This separation avoids mixing application-level packaged performance with fixture-corrected quantities.

**Fabrication repeatability**

A newly fabricated resonator using the same Shape-5 boundary concept was characterized independently. The test is intended to verify that the characteristic main resonance and reduced lateral-mode structure are not unique to a single specimen.

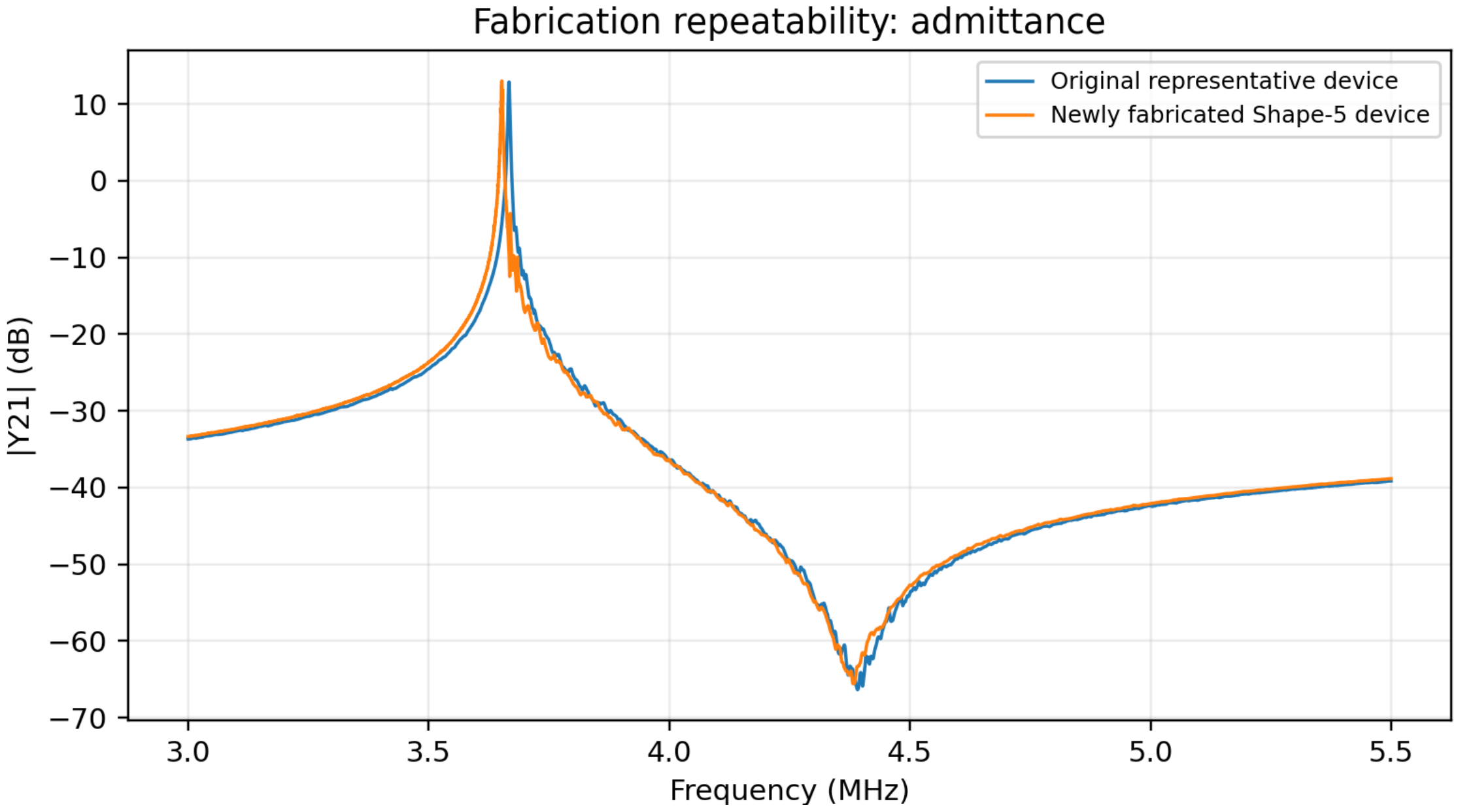

***Fig. S13.*** *Admittance comparison between the original representative device and a newly fabricated device.*

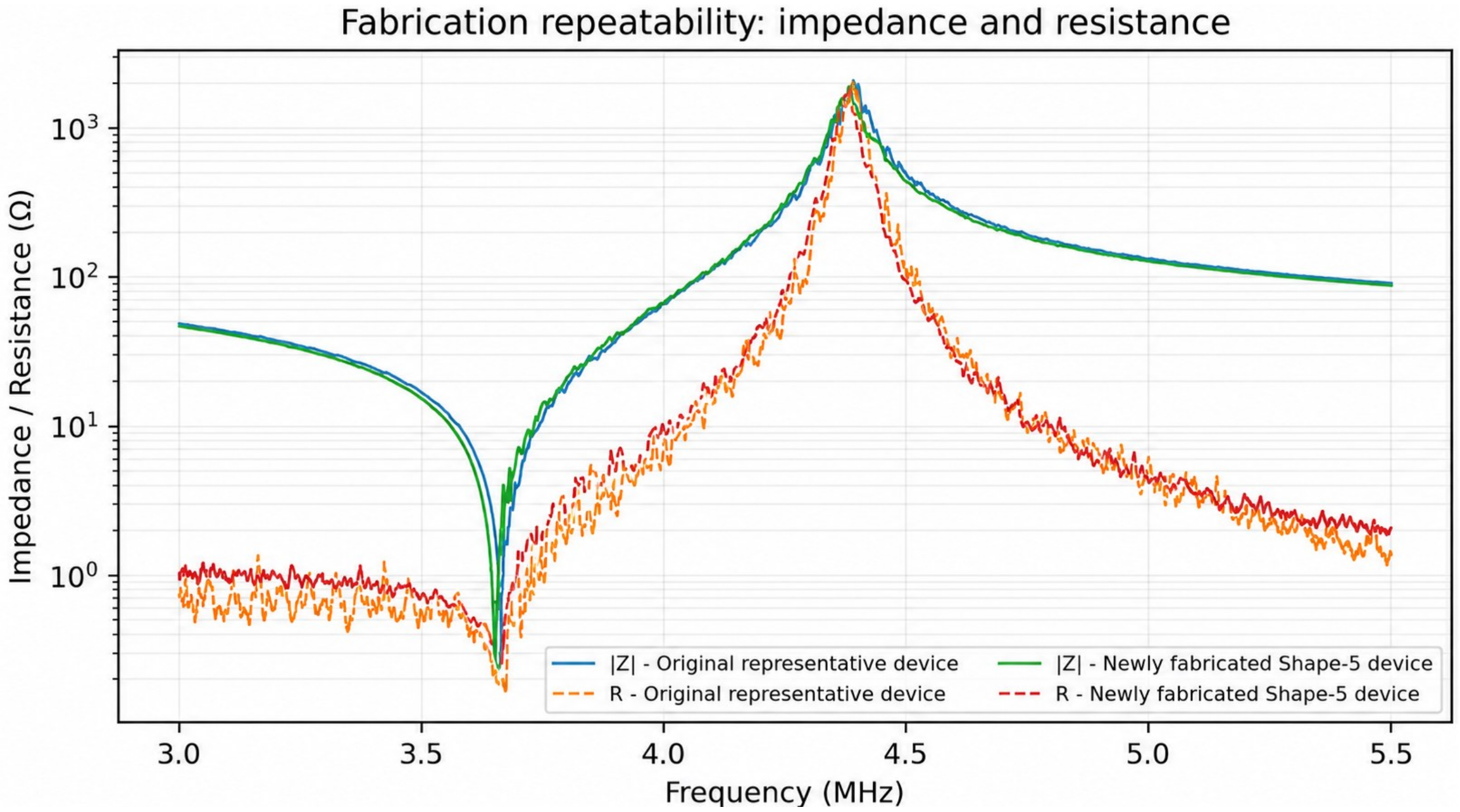

***Fig. S14.*** *Impedance and resistance comparison between the original device and a newly fabricated device.*

| Device | $f_s$ (MHz) | $f_p$ (MHz) | $k_t^2$ (%) | 3-dB $Q_s$ |
|---|---|---|---|---|
| Original representative device | 3.67 | 4.39 | 53.57 | 1146 |
| Newly fabricated Shape-5 device | 3.65 | 4.38 | 54.30 | 1259 |

***Table S8.*** *Measured resonator parameters for fabrication repeatability of the same Shape-5 design.*

The new device fabricated using an identical design reproduces the same qualitative resonance topology and a similar impedance/resistance trajectory. The modest frequency and $Q$ differences are consistent with fabrication, attachment, and interconnection variation and are acknowledged as part of the device-to-device uncertainty.

**VNA source-level dependence**

The packaged resonator was remeasured over a source-level sweep from −15 to +20 dBm using the same VNA configuration. The available archived files correspond to −15, −5, 0, +10, and +20 dBm. These measurements are an excitation-level sensitivity test and should not be interpreted as a full converter-level large-signal power-handling demonstration, because the net acoustic power dissipated in the resonator is not independently measured.

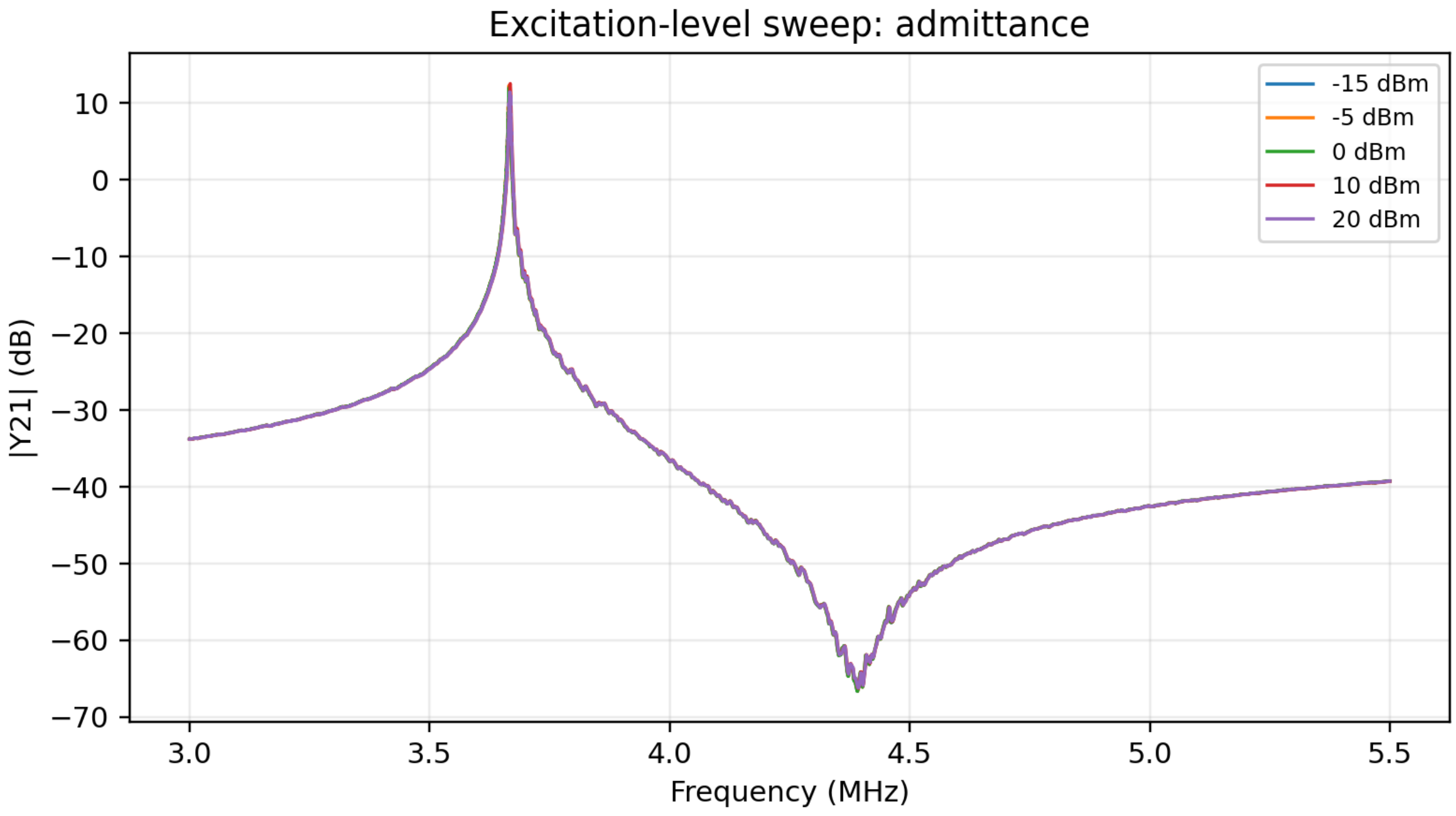

***Fig. S15***. *Admittance measured at source levels from −15 to +20 dBm.*

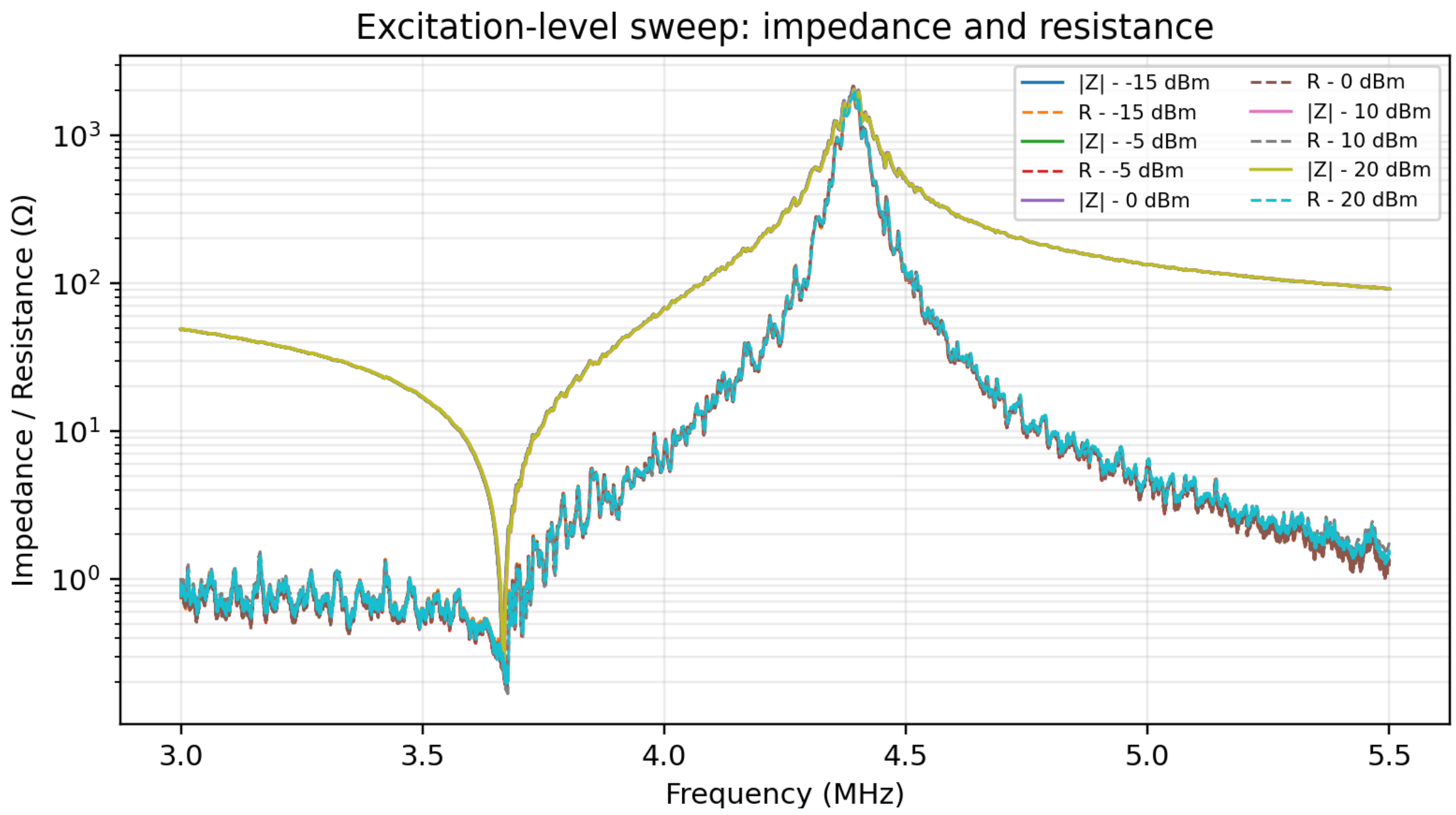

***Fig. S16.*** *Impedance magnitude and resistance measured at source levels from −15 to +20 dBm.*

| **Source level** | $f_s$ **(MHz)** | $f_p$ **(MHz)** | **3-dB** $Q_s$ |
|---|---|---|---|
| -15 dBm | 3.67 | 4.39 | 1183 |
| -5 dBm | 3.67 | 4.39 | 1111 |
| 0 dBm | 3.67 | 4.39 | 1111 |
| 10 dBm | 3.67 | 4.39 | 1079 |
| 20 dBm | 3.67 | 4.39 | 991 |

***Table S9.*** *Measured resonance parameters and 3-dB series quality factor of the same physical resonator under different source excitation levels.*

Using the same MATLAB $Y_{21}$ extraction, $Q_s$ ranges from 991 to 1183. The device temperature was monitored during the measurements using an infrared temperature gun and remained approximately 24 °C throughout the sweep. No large monotonic resonance shift is observed over this range, up to the highest applied VNA source level of +20 dBm (100 mW nominal source power).

**Quality-factor extraction and numerical differentiation**

Three quality-factor quantities are distinguished in the revised manuscript: (i) $Q_s$ from the 3-dB bandwidth around the series resonance, (ii) Bode $Q$ evaluated at $f_s$, and (iii) the maximum local Bode $Q$ near resonance. These values need not be identical because they weight the resonance line shape, phase slope, and electrical loss differently. The 3-dB $Q$ is used as the primary series-resonance loss metric; Bode $Q$ at fs provides a phase-slope-based cross-check, and the maximum Bode $Q$ is reported only as a local peak value, not as a replacement for $Q_s$. It is also utilized to compare with prior art that only reports the maximum Bode $Q$.

For the validation data reported here, $Q_s$ is recalculated exactly from the author-supplied MATLAB routine: S parameters are converted to Y parameters using s2y(S,50), $Y_{21}$ is expressed in dB, $f_s$ is the $Y_{21}$ maximum, and the 3-dB bandwidth is the span between the first and last sampled frequencies for which $Y_{21}$ remains within 3 dB of its maximum. No interpolation is introduced in these recalculated $Q_s$ values.

For Bode-$Q$ extraction, the phase of the measured $S_{11}$ was first unwrapped to remove 2π phase discontinuities. The phase derivative with respect to angular frequency was then calculated numerically using MATLAB's gradient function. The Bode $Q$ was calculated from Eq. (4) in the main manuscript. The $f_s$ and $f_p$ were identified from the maximum and minimum, respectively, of the magnitude of $Y_{11}$, where $Y_{11} = [(1 - S_{11})/(1 + S_{11})]/Z_0$ with $Z_0 = 50\ \Omega$. The Bode-$Q$ values at $f_s$ and $f_p$ were obtained directly from the corresponding points on the calculated $Q_B$ curve. The maximum Bode $Q$, $Q_{B(\mathrm{max})}$, was determined as the maximum absolute value of $Q_B$ within the frequency interval between $f_s$ and $f_p$. The same extraction procedure was applied consistently to all measurements. Because $Q_{B(\mathrm{max})}$ is determined from a local maximum of a numerically differentiated phase response, it is inherently more sensitive to point-to-point measurement noise than $Q_{B(fs)}$. We therefore report both $Q_{B(fs)}$ and $Q_{B(\mathrm{max})}$ and distinguish the two quantities explicitly.

For the representative device, the manuscript reports $Q_s$ = 1146, $Q_{B(fs)}$ = 1621, and $Q_{B(\mathrm{max})}$ = 2896. The separation among these values reflects both the different $Q$ definitions and the sensitivity of the local derivative maximum to the detailed line shape. The principal conclusions are therefore not based solely on $Q_{B(\mathrm{max})}$.

## Quantification of Spurious-Mode Suppression Using Valley-to-Peak Residual Energy

### Main focus

Spurious modes appear as narrow resistance fluctuations superimposed on the broad main valley-to-peak response of a piezoelectric resonator. Direct visual comparison is useful but subjective, especially when different devices operate at different frequencies. The analysis below converts each resistance spectrum to log scale, extracts only the main valley-to-peak region, fits a smooth one-valley-to-one-peak main-mode curve, and quantifies the residual fluctuations around that curve.

The method is intended to compare the smoothness of resistance spectra, not to refit the full electromechanical equivalent circuit. Equivalent-network descriptions of piezoelectric resonators provide the physical motivation that the principal resonant response should be smooth at the main-mode level[2], while narrow residual features are treated as spurious-mode contributions.

### Input data and preprocessing

The resistance is analyzed on a logarithmic scale:

$$y_i = \log_{10}(R_i) \tag{S4}$$

The log transformation prevents high resistance regions from dominating the residual calculation solely because the resistance spans several orders of magnitude.

### Valley-to-peak analysis window

For each device, the user specifies a frequency window around the main resistance valley and a second window around the main resistance peak. The code identifies the lowest log-resistance point in the valley window and the highest log-resistance point in the peak window. Only the interval between these two detected frequencies is analyzed:

$$x_i = \frac{f_i - f_{valley}}{f_{peak} - f_{valley}} \tag{S5}$$

This normalization maps the valley to 0 and the peak to 1, allowing residual curves from devices with different operating frequencies to be compared on the same horizontal scale.

For reproducibility, the exact search windows used in the archived analysis are listed in Table S10. These windows only identify the principal valley and peak; they are not themselves used as the fitting interval. The default implementation uses 11 control points, the $50^{th}$ percentile (local median) within each interior control bin, a three-point median smoothing of the control values, monotonic-rise enforcement, and no endpoint exclusion.

| Device | Valley search window (MHz) | Peak search window (MHz) | Detected valley (MHz) | Detected peak (MHz) |
|---|---|---|---|---|
| TS-LN (Our Work) | 3.55–3.72 | 4.25–4.45 | 3.67450 | 4.39110 |
| TE-LN[4] | 6.70–6.95 | 7.50–7.75 | 6.81862 | 7.68307 |
| TS-LN[5] | 5.85–6.15 | 6.70–7.25 | 5.97808 | 6.78728 |
| TE-LN[6] | 10.00–10.30 | 11.20–11.38 | 10.24135 | 11.29228 |
| TE-LT[7] | 6.20–6.32 | 6.43–6.52 | 6.24407 | 6.47336 |
| TE-LN[8] | 19.05–19.35 | 21.25–21.50 | 19.29429 | 21.38327 |
| TS-LN[9] | 3.55–3.75 | 4.20–4.45 | 3.63250 | 4.34400 |

***Table S10.*** *Explicit valley and peak search windows used for the resistance-residual analysis, together with the extrema detected from the archived traces.*

**Smooth main-mode fitting curve**

The smooth fitting curve is designed to retain only the broad main-mode evolution from the valley to the peak. It should not follow the narrow spurious ripples. The implemented procedure is:

1. Generate evenly spaced control frequencies between the detected valley and the detected peak.

2. Fix the first control point at the detected valley and the last control point at the detected peak.

3. For each interior control-frequency bin, assign the local median of $\log_{10}(R)$ as the control value. The median reduces the influence of narrow upward or downward spurious features.

4. Apply light median smoothing to the interior control values while preserving the valley and peak endpoints.

5. Enforce a monotonic rise from valley to peak so that the fitted main-mode curve contains only one valley and one peak.

6. Interpolate the control points using shape-preserving piecewise cubic Hermite interpolation (PCHIP). This choice avoids artificial oscillations when the control points are monotonic[3].

The numerical fitting curve used for all reported values is therefore defined directly by the PCHIP interpolation of the monotonic control-point set:

$$\hat{y}(f) = \mathrm{PCHIP}(\{(f_{\mathrm{k}}, \tilde{y}_{\mathrm{k}})\}_{k=1}^{N_{ctrl}}; f) \tag{S6}$$

Here, $f_k$ are the evenly spaced control frequencies and $\tilde{y}_k$ are the median-based, lightly smoothed, monotonic control values described above. Equation S6 is thus a compact notation for the implemented interpolation procedure; no separate analytic fitting function is used to generate the reported residuals.

Median/percentile-based control-point selection is a robust estimation strategy; it limits the influence of local outliers or narrow spurious peaks on the estimated smooth trend[4].

**Residual definition**

The spurious residual is defined as the difference between the measured log-resistance and the fitted smooth main-mode curve:

$$e_i = y(x) - \hat{y}(x) = \log_{10}(R_i) - \log_{10}(R_{fit,i}) \tag{S7}$$

All three metrics are calculated only from data points inside the valley-to-peak interval. Lower values indicate a smoother valley-to-peak response and smaller residual fluctuations relative to the smooth main-mode baseline.

**Overall spurious strength**

The overall spurious strength is the root-mean-square residual:

$$S_{overall} = \sqrt{\frac{1}{N}\sum_{i=1}^{N} e_i^2} \tag{S8}$$

This metric measures the average residual fluctuation energy in $\log_{10}(R)$. It is the primary smoothness metric.

**95th-percentile residual amplitude**

As a complementary peak-oriented metric, the analysis also reports the 95th percentile of $|e(f)|$ within the same valley-to-peak interval. This statistic emphasizes stronger local fluctuations while being less sensitive than a single maximum to isolated digitization or measurement-noise points.

**Residual per MHz**

The residual per MHz integrates the squared residual over frequency and normalizes by the analyzed frequency span:

$$E_{rs} = \text{Residual per MHz} = \frac{1}{f_{peak} - f_{valley}} \int_{f_{valley}}^{f_{peak}} e(f)^2 \, df \tag{S9}$$

For discrete data, the trapezoidal numerical integration used in the MATLAB code is:

$$E_{rs} = \frac{trapz(f_i, e_i^2)}{f_{peak} - f_{valley}} \tag{S10}$$

This metric is particularly useful when different devices contain different numbers of sampling points or slightly different valley-to-peak spans. The frequency normalization makes the value comparable as residual energy density rather than as total unnormalized residual energy.

**MATLAB results**

We used MATLAB to process all the data. The resistance spectra for previously reported devices were extracted from published reports as shown in Fig. S17. Fig. S17 shows the measured resistance $R$ plotted on a logarithmic vertical scale, and the frequency is given in MHz. The panels compare the present apodized TS-mode $LiNbO_3$ resonator with previously reported resonator spectra digitized or extracted for benchmarking.

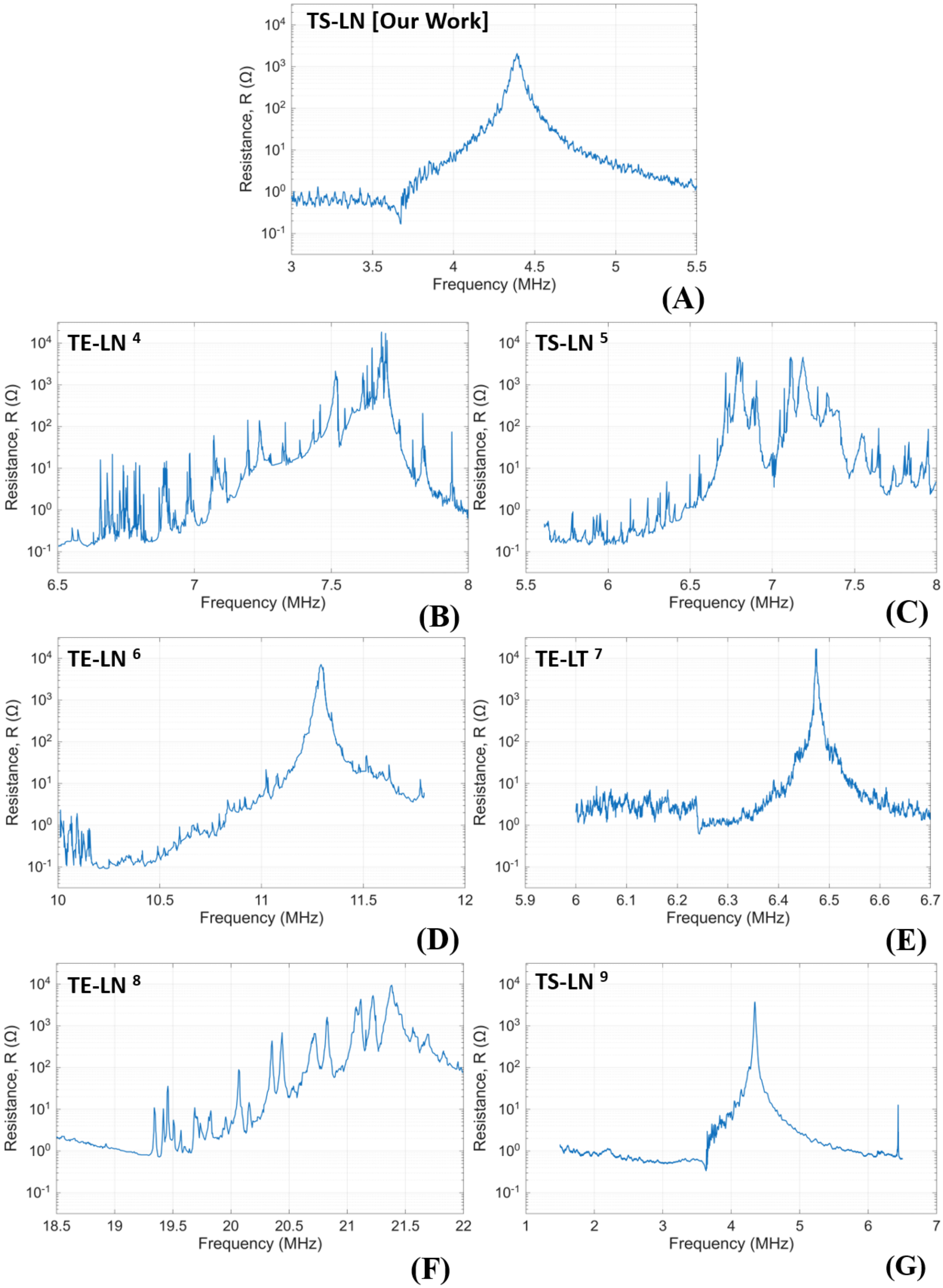

***Fig. S17.*** *Resistance–frequency spectra used for spurious-mode benchmarking. The spectra are plotted on a logarithmic R scale to compare the main resonance response and spurious-mode features. Rows correspond to the present work, TE-LN* [5]*, TS-LN* [6]*, TE-LN* [7]*, TE-LT* [8]*, TE-LN* [9]*, and TS-LN* [10]*.*

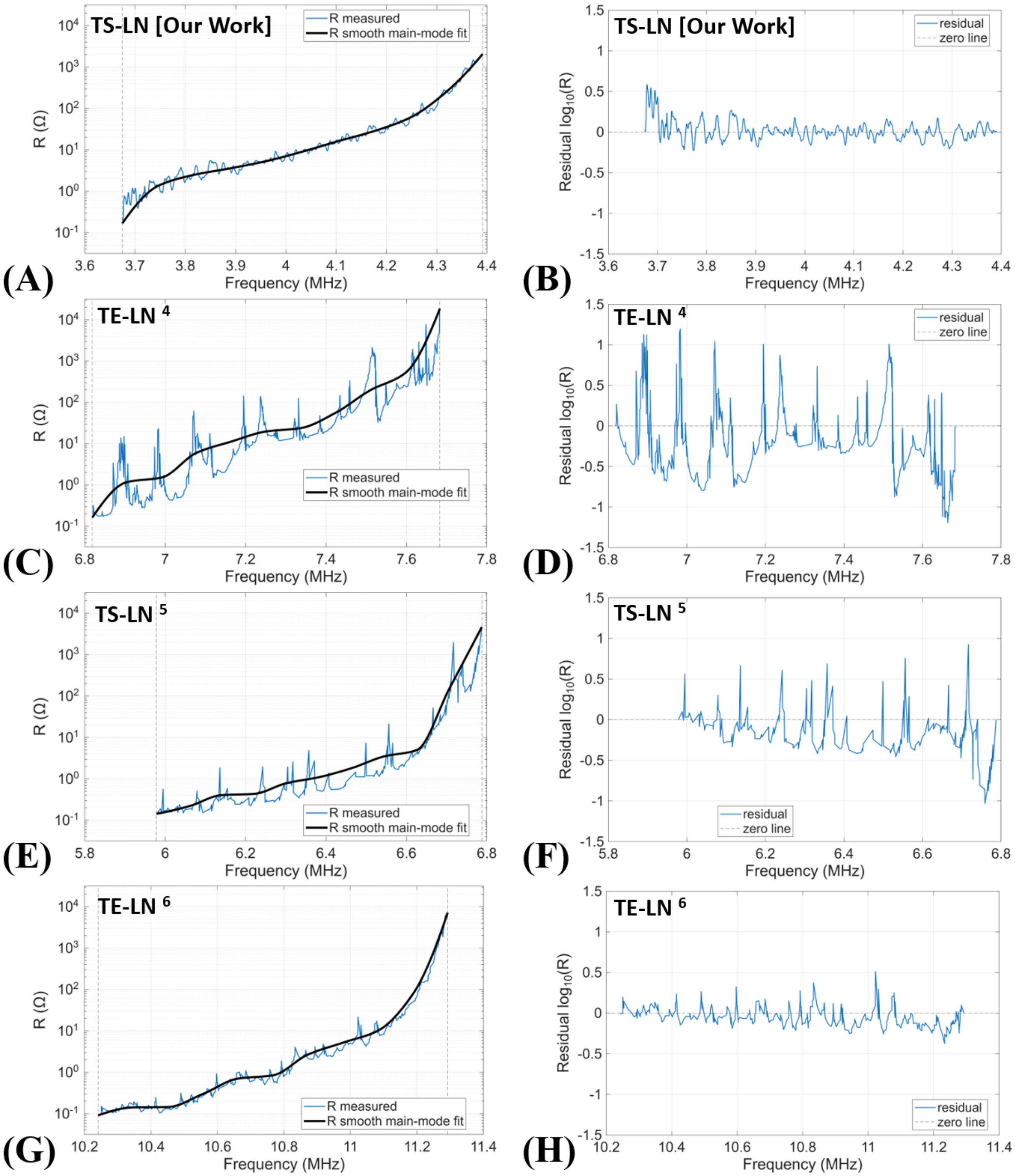
TS-LN [Our Work]
R measured
R smooth main-mode fit
R (Ω)
Frequency (MHz)
(A)
TS-LN [Our Work]
residual
zero line
Residual log10(R)
Frequency (MHz)
(B)
TE-LN 4
R measured
R smooth main-mode fit
R (Ω)
Frequency (MHz)
(C)
TE-LN 4
residual
zero line
Residual log10(R)
Frequency (MHz)
(D)
TS-LN 5
R measured
R smooth main-mode fit
R (Ω)
Frequency (MHz)
(E)
TS-LN 5
residual
zero line
Residual log10(R)
Frequency (MHz)
(F)
TE-LN 6
R measured
R smooth main-mode fit
R (Ω)
Frequency (MHz)
(G)
TE-LN 6
residual
zero line
Residual log10(R)
Frequency (MHz)
(H)

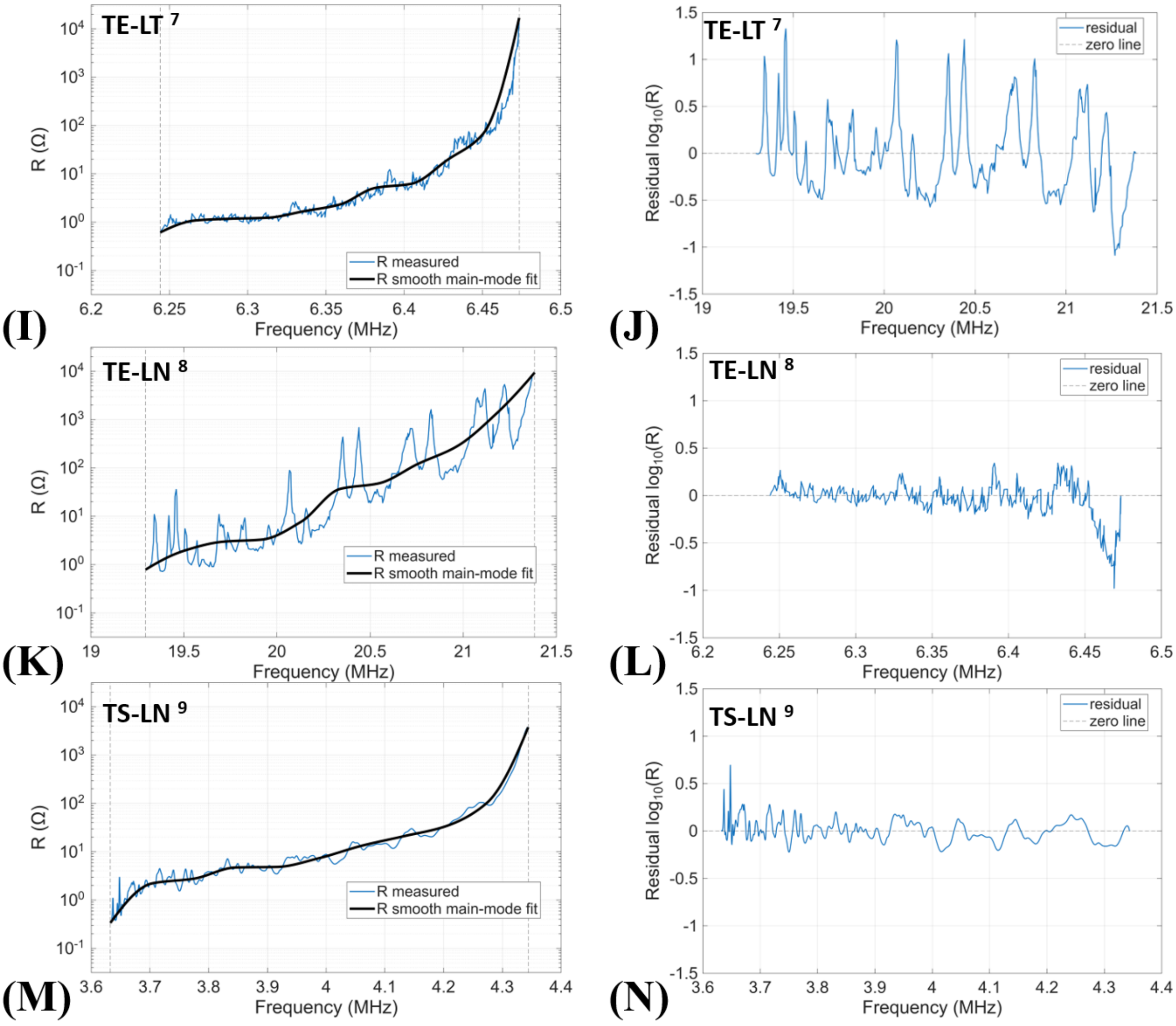

***Fig. S18.*** *Valley-to-peak fitting and residual extraction used for spurious-mode benchmarking. For each resonator, the left panel shows the measured resistance spectrum and the smooth main-mode fit over the valley-to-peak interval, and the right panel shows the corresponding residual, defined as the difference between the measured and fitted log-resistance. The dashed vertical lines indicate the detected valley and peak boundaries used for the analysis. From top to bottom, the rows correspond to the present TS-LN resonator, TE-LN [5], TS-LN [6], TE-LN [7], TE-LT [8], TE-LN [9], and TS-LN [10].*

Fig. S18 presents the left-column panels, which are the measured resistance and the smooth one-valley-to-one-peak main-mode fit. The right-column panels show that the corresponding residual equation: $e(f) = \log_{10}(R) - \log_{10}(R_{fit})$. The vertical dashed lines identify the detected valley and peak boundaries used for the metric calculation. The smooth fit is constrained not to follow narrow ripples; therefore, those ripples remain in the residual.

Fig. S19(A) shows that the metric is the RMS value of the residual in $\log_{10}(R)$. Lower values indicate a smoother measured curve relative to the smooth main-mode trend. Fig. S19(B) integrates the squared residual over the analyzed frequency interval and normalizes by the interval width. Lower values indicate lower total spurious fluctuation energy per unit frequency.

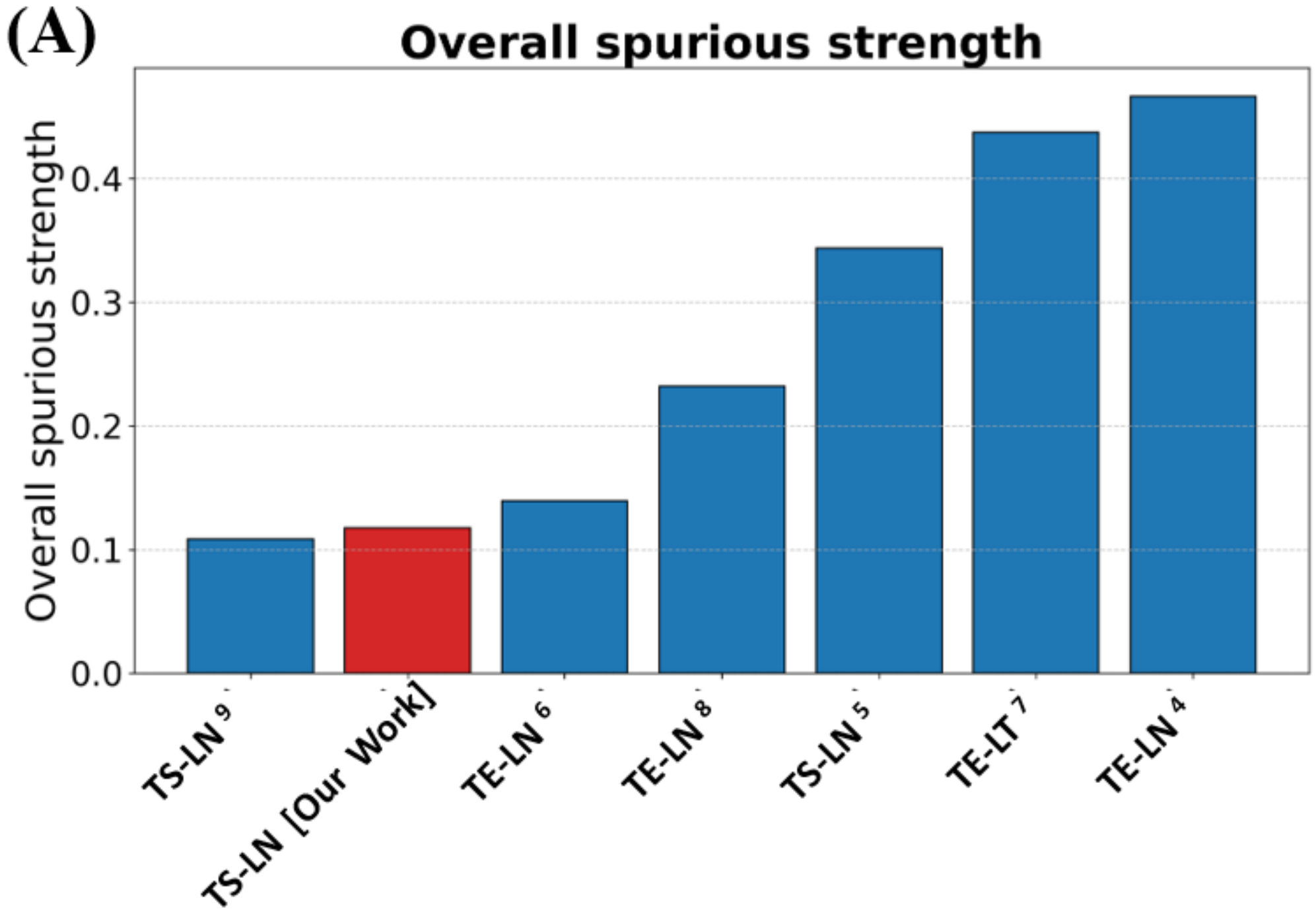

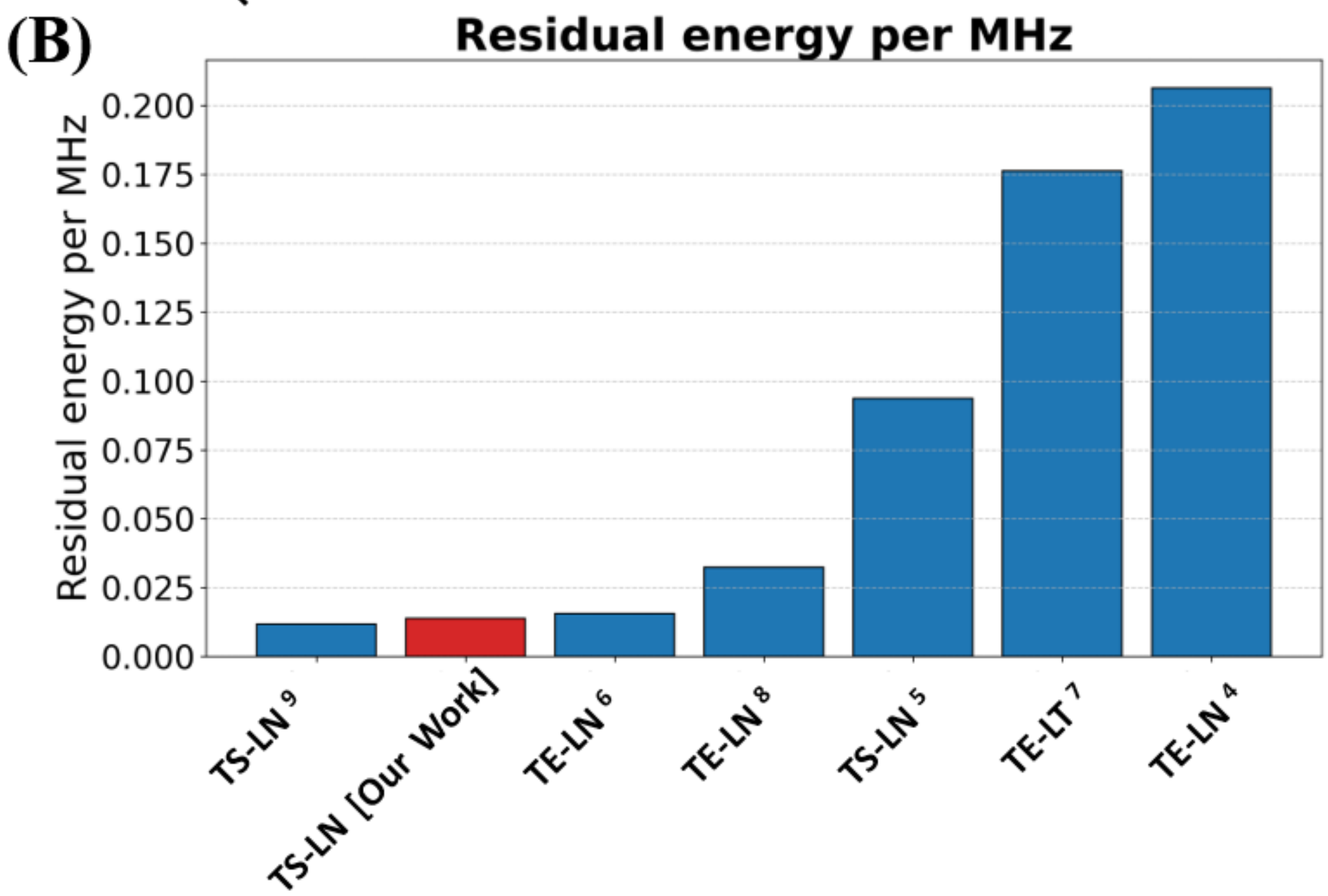

***Fig. S19.*** *(A) Overall spurious strength extracted from the valley-to-peak residual for all devices. (B) Residual per MHz extracted from the valley-to-peak residual for all devices.*

**Robustness and cross-validation of the spurious-response assessment**

The residual metric is used as a comparative smoothness measure rather than an absolute definition of a spurious-free resonator. To evaluate sensitivity to the empirical baseline, the archived resistance traces were reprocessed using nine settings combining 9, 11, and 13 control points with 40th, 50th, and 60th percentile control values. Table S11 summarizes the default metrics and the resulting residual-per-MHz ranges.

| Device | RMS residual | 95th \|residual\| | Residual/MHz (default) | Residual/MHz range (9 settings) |
|---|---|---|---|---|
| TS-LN (Our Work) | 0.1175 | 0.2167 | 0.0138 | 0.0128–0.0191 |
| TE-LN[4] | 0.4664 | 0.9188 | 0.2062 | 0.1639–0.2682 |
| TS-LN[5] | 0.3439 | 0.7172 | 0.0937 | 0.0669–0.1551 |
| TE-LN[6] | 0.1395 | 0.2626 | 0.0157 | 0.0094–0.0288 |
| TE-LT[7] | 0.2323 | 0.5984 | 0.0324 | 0.0188–0.0544 |
| TE-LN[8] | 0.4375 | 0.9243 | 0.1762 | 0.1528–0.2155 |
| TS-LN[9] | 0.1085 | 0.1997 | 0.0118 | 0.0099–0.0192 |

***Table S11.*** *Residual-based metrics for the default analysis and sensitivity of residual energy per MHz to nine baseline settings (9/11/13 control points × 40th/50th/60th percentile control values).*

For the present TS-LN resonator, residual energy per MHz remains within 0.0128–0.0191 across all nine settings, and the device remains among the three lowest-residual traces. The Spearman rank correlation with the default ordering ranges from 0.86 to 1.00, indicating that the relative comparison is robust to reasonable baseline variations. The user-defined valley/peak windows remain an empirical element of the procedure and are therefore reported explicitly in Table S10 for reproducibility.

As an independent electrical cross-check, the uninterrupted inductive bandwidth is evaluated directly from $Im(Z_{series})>0$ and does not depend on residual fitting. For the representative 560 mm² Shape-5 resonator, the inductive interval extends from 3.667 to 4.390 MHz, corresponding to approximately 723 kHz; values across the area-scaled series range from 513 to 723 kHz (Table S5).